\documentstyle[epsf,12pt,aasms4,psfig]{article}
\begin{document}

  

\def\gsim{\;\lower.6ex\hbox{$\sim$}\kern-7.75pt\raise.65ex\hbox{$>$}\;}
\def\lsim{\;\lower.6ex\hbox{$\sim$}\kern-7.75pt\raise.65ex\hbox{$<$}\;}
\def\aa{A\&A }
\def\mb{$m_{F439}$}
\def\mv{$m_{F555}$}
\def\mgu{$m_{F380}$}
\def\mbv{$m_{F439}-m_{F555}$}
\def\secpoint{$\arcsec$.}
\def\sigmada{$\sigma_{DAO}$}
\newcommand{\TEFF}{\mbox{$T_{\rm eff}$}}
\newcommand{\TSTAR}{\mbox{T$_{\ast}$}}
\newcommand{\TE}{\mbox{T$_e$}}
\newcommand{\NE}{\mbox{n$_e$}}
\newcommand{\HII}{\mbox{H~II}}
\newcommand{\Ms}{M$_{\odot}$}
\newcommand{\Myr}{M$_{\odot}$~yr$^{-1}$}
\newcommand{\etal}{\mbox{et al.}}
\newcommand{\MSUN}{\mbox{M$_{\odot}$}}
\newcommand{\Halpha}{\mbox{H$_{\alpha}$}}
\newcommand{\OII}{\mbox{[OII]}}
\newcommand{\OIII}{\mbox{[OIII]}}
\newcommand{\SII}{\mbox{[SII]}}
\newcommand{\NII}{\mbox{[NII]}}
\newcommand{\Mgtw}{\mbox{Mg$_2$~}}
\newcommand{\Hbeta}{\mbox{H$\beta$}}
\newcommand{\Fetw}{\mbox{Fe52~}}
\newcommand{\Feth}{\mbox{Fe53~}}
\newcommand{\Lsun}{\mbox{L$_{\odot}$}}
\newcommand{\Zsun}{\mbox{Z$_{\odot}$}}
\newcommand{\BCvth}{\mbox{$BC_{V,th}$}}

\title{The Resolved Stellar Population of the Post-Starburst Galaxy
NGC~1569\footnote{Based on observations with the NASA/ESA Hubble
Space Telescope, obtained at the Space Telescope Science Institute,
which is operated by AURA for NASA under contract NAS5-26555}} 

\vspace{0.5in}

\author{Laura Greggio$^{2,3}$, Monica Tosi$^4$, Mark Clampin$^5$, 
        Guido De Marchi$^6$, Claus Leitherer$^5$, Antonella Nota$^{5,7}$, 
        and Marco Sirianni$^8$}

\vspace{0.25in}

\affil{$^2$ Universit\`a di Bologna, Dipartmento di Astronomia, 
       Via Zamboni 33, I-40126 Bologna, Italy\\
       e-mail: greggio@astbo3.bo.astro.it}

\affil{$^3$ Universitaets Sternwarte Muenchen, 
       Scheinerstrasse 1, D-81679 Muenchen, Germany\\
       e-mail: greggio@usm.uni-muenchen.de}

\affil{$^4$ Osservatorio Astronomico di Bologna, Via Zamboni 33,
       I-40126 Bologna, Italy\\
       e-mail: tosi@astbo3.bo.astro.it}

\affil{$^5$ Space Telescope Science Institute, 
       3700 San Martin Drive, Baltimore, MD 21218\\
       e-mail: clampin@stsci.edu, leitherer@stsci.edu, nota@stsci.edu}

\affil{$^6$ European Southern Observatory, 
        Karl-Schwarzschild-Str. 2, 
        D-85748 Garching, Germany\\
        e-mail: demarchi@eso.org}

\affil{$^7$ Affiliated with the Astrophysics Division, Space
       Science Department of the European Space Agency} 

\affil{$^8$ Universit\`a di Padova, Dipartmento di Astronomia, 
       Vicolo dell'Osservatorio 5, I-40100 Padova, Italy\\
       e-mail: sirianni@astrpd.pd.astro.it}

\begin{abstract}

We present WFPC2--HST photometry of the resolved stellar population in the
post-starburst galaxy NGC~1569. The color-magnitude diagram (CMD)
derived in the F439W and F555W photometric bands contains $\sim$2800
stars with photometric error $\leq$ 0.2 mag down to \mb, \mv~$\simeq$~26,
and is complete for \mv~$\lsim$~23. Adopting the literature
distance modulus and reddening, our CMD samples stars more massive
than $\sim$4~\MSUN, allowing us to study the star
formation (SF) history over the last $\sim$0.15~Gyr.
The data are interpreted using
theoretical simulations based on stellar evolutionary models. The synthetic
diagrams include photometric errors and incompleteness
factors. Testing various sets of tracks, we find that the ability of
the models to reproduce the observed features in the CMD is strictly
related to the shape of the blue loops of the sequences with  
masses around 5~\MSUN. The field of NGC~1569
has experienced a global burst of star formation of duration $\gsim$0.1~Gyr,
ending $\sim$ 5$-$10~Myr ago. During the burst, the SF rate was
approximately constant, and, if quiescent periods occurred, they
lasted less than $\sim$10~Myr. The level of the SF rate
was very high: for a single-slope initial mass function (IMF) ranging 
from 0.1 to 120~\MSUN\ we
find values of 3, 1, and 0.5~\Myr\ for $\alpha = 3$, 2.6, and 2.35 (Salpeter),
respectively. When scaled for the surveyed
area, these rates are approximately 100 times larger than found in the most
active dwarf irregulars in the Local Group. The data are consistent
with a Salpeter IMF, though our best models indicate slightly
steeper exponents.
We discuss the implications of our results in the general context of 
the evolution of dwarf galaxies.

\end{abstract}

 
\section{Introduction}

Understanding the status and evolution of dwarf galaxies is of crucial
importance in current astrophysics and cosmology. 
They are important ingredients in common scenarios
of galaxy formation, either as building blocks or as left-overs of the
formation process (Silk 1987). In addition, dwarf galaxies forming at
redshift around and below 1 have been suggested to be responsible for
the excess of faint blue galaxies seen in deep photometric surveys
(Babul $\&$ Rees 1992; Babul $\&$ Ferguson 1996). 

Dwarf galaxies come in three main classes: (i) dwarf spheroidals,
which contain population~II stars and have low surface brightness (e.g., 
Wirth \& Gallagher 1984); (ii) dwarf irregulars, with various levels
of star-formation activity and a high gas mass (Hunter \& Gallagher 1986);
(iii) starbursting dwarfs, including blue compact dwarfs (Thuan 1991), H~II
galaxies (Terlevich \etal\ 1991), and blue amorphous galaxies (Sandage \&
Brucato 1979). The boundary between dwarf and giant galaxies is usually 
taken to be around $M_{\rm{B}} = -16$ mag (Tammann 1994).

An intriguing property of dwarf galaxies is the pronounced dichotomy between
the two main classes of dwarf spheroidals and dwarf irregulars: most dwarfs
are either gas-poor spheroids or gas-rich disk-like systems, with very few
transition objects (van den Bergh 1977). The relationship between the two
classes is still poorly understood: attempts to unify the complex zoo of
dwarf galaxies have not been fully successful (Binggeli 1994). Some effort
has been made to understand the interrelationship between individual classes
in terms of an evolutionary sequence (e.g., Gallagher, Hunter, \& Tutukov
1984). 
Starbursting dwarfs can play a key role: a powerful starburst might
strip a gas-rich dwarf irregular of its gas and transform it into a 
gas-poor dwarf-elliptical (Kormendy 1985; Marlowe 1997).

Understanding the star-formation histories in starbursting dwarf galaxies is 
a prerequisite for understanding evolutionary connections between different 
object classes (e.g., Gallagher 1996) since during cosmologically brief periods 
of time their star formation increases 
dramatically. This has profound implications for the interstellar medium. 
Stellar winds and supernovae inject energy and may cause a `blow-out', 
initiating a galactic superwind (Heckman 1995). Numerous examples
of galactic superwinds are known from observations (e.g., Heckman, Armus,
\& Miley 1990) and their cosmological implications have been discussed (e.g.,
De Young \& Heckman 1994). However, attempts to relate the superwind 
properties and the star-formation histories in individual galaxies are quite
rare and do not go beyond a qualitative level (Heckman \etal\ 1990; Leitherer,
Robert, \& Drissen 1992). Yet establishing such a relation is crucial for
theoretical models of the superwind hydrodynamics (Suchkov \etal\ 1994;
Tenorio-Tagle \& Mu\~noz-Tu\~non 1997).

NGC~1569 (= UGC~3056 = Arp~210 = VII~Zw~16 = IRAS~4260+6444) has been
indicated by Gallagher, Hunter, $\&$ Tutukov (1984) as an outstanding
object in their sample of active star forming galaxies. Indeed, it is a
prime candidate for a detailed study of the star-formation history in 
a starburst galaxy since at $D = 2.2 \pm 0.6$~Mpc it is the closest starburst
galaxy known (Israel 1988a). Adopting the literature reddening
$E(B-V)$ = 0.56 (Israel 1988a), photometry of individual stars is
feasible with HST down to $M_{V,0}\simeq -1.5$. This corresponds
to stars with mass of about 3~\Ms\ in their core-helium burning phase,
and consequently to a look-back time of $\sim$0.4~Gyr
(e.g., Schaller \etal\ 1992).

The properties of NGC~1569 are typical of a dwarf galaxy. With a
distance modulus of $(m-M)_0$ = 26.71, its total absolute B magnitude
is $M_{B.0} \sim -17$ (Israel 1988a), which is intermediate between
the two Magellanic Clouds. 
Its total mass and hydrogen content are estimated $M \simeq 3.3 \times
10^8$~\Ms\ and $M_H \simeq 1.3 \times 10^8$~\Ms, respectively (Israel 1988a).
Numerous studies of the chemical composition exist. The published range of
oxygen abundances is very narrow: $12 + \log (O/H) = 8.25$ (Hunter, Gallagher,
\& Rautenkranz 1982), 8.37 (Calzetti, Kinney, \& Storchi-Bergmann 1994),
8.26 (Devost, Roy, \& Drissen 1997), 8.29 (Gonz\'alez-Delgado et al. 1997),
and 8.19 (Kobulnicky \& Skillman 1997; Martin 1997). Searches for chemical
composition gradients were done by Devost et al. and Kobulnicky \& Skillman.
No significant evidence for chemical inhomogeneities was found. If [O/Fe]~=~0.0
is assumed, the average metallicity of NGC~1569 is $Z \simeq 0.25$~\Zsun, with
about 0.2~dex uncertainty. Thus, NGC~1569 is a gas-rich system with SMC-like
composition in a relatively early stage of its chemical evolution.

The most detailed study of the warm ($\sim$10$^4$~K) gas of NGC~1569 was done 
by Waller (1991), who found evidence for a link between the star-formation
history and the morphology and kinematics of extended gas structures. The
same connection is suggested by hot ($\sim$10$^7$~K) gas observed with
the ROSAT and ASCA satellites (Heckman \etal\ 1995; Della Ceca \etal\ 1996). 
The extended X-ray emission is consistent with a starburst driven 
galactic superwind which could in principle lead to a large-scale disruption 
of the interstellar medium. Evidence for large numbers of supernovae was also
found from the non-thermal radio spectrum (Israel 1988b).

One of the most spectacular properties of NGC~1569 are its `super star 
clusters'. These high-density star clusters were first detected and discussed
by Arp \& Sandage (1985) and Melnick \& Moles, \& Terlevich (1985) and later
studied by O'Connell, Gallagher, \& Hunter (1994), Ho \& Filippenko
(1996), De~Marchi \etal\ (1997), and Gonz\'alez-Delgado \etal\ (1997). The
super star clusters reflect a recent starburst in which at least $10^5$~\Ms\
of gas was transformed into stars within about 1~pc. Super star clusters are
similar to young Galactic globular clusters observed shortly after their
formation (Meurer 1995).

Pre-Costar HST studies of the field population were done by O'Connell \etal\
(1994) and Vallenari \& Bomans (1996) (hereinafter VB). 
According to the latter authors, NGC~1569 
experienced a strong starburst until a few Myr ago, when its activity
subsided. The repaired HST offers the possibility of significant improvement
over these previous studies. We have therefore embarked on an extensive
spectroscopic and photometric HST study in an attempt to confront the 
observational data with the predictions of theoretical models, and derive
detailed informations on the star formation (SF) history in NGC~1569. 
In a first paper (De Marchi \etal\ 1997) we discussed the photometry
of the SSCs, while in this paper we present the results for 
the resolved stars in the same field, which are interpreted in
terms of the recent SF history via theoretical simulations.
The data and their
analysis are described in Section~2. The derived color-magnitude diagram and 
luminosity function are in Section~3. Our theoretical models are described in
Section~4. They are compared with the observations in Section~5. 
In Section~6 we discuss our results and their implications for the
evolution of dwarf galaxies, and we present in Section~7 our general
conclusions.
   
\section{Data reduction and analysis}

The observations of NGC\,1569 described in this paper were obtained
with WFPC2 on board the HST on 1996 January
10 (UT). NGC\,1569 was centered in the field of view of the planetary
camera (PC) yielding an effective plate scale of
$0\secpoint046$\,pixel$^{-1}$ and a $35\arcsec \times 35\arcsec$ field
of view. The plate scale and field of view of the adjacent wide field
chips (WF) are $0\secpoint1$\,pixel$^{-1}$ and $80\arcsec \times
80\arcsec$, respectively. Images were obtained in three broadband
filters, F380W, F439W and F555W, whose photometric properties are
similar --- yet not identical --- to the Johnson $U, B, V$ filters, and
are described in detail in the WFPC2 handbook (Biretta 1996).  Since
the photometric quality of the F380W images turned out to be poor, it
was not possible to measure stars fainter than $\sim$24~mag
with photometric errors smaller than $\sim$0.2~mag. We have then
concentrated our analysis on the F439W and F555W data only, and used
the F380W images to check for spurious detections (see below).

A set of 16 deep exposures was collected for each of these filters, all
of duration 630 s in F439W and 450 s in F555W. Images were taken at 4
different positions, dithered by a few pixel in order to improve the
sampling of the final combined images, to recover faint stars with
reliability, and to eliminate high frequency noise.  The whole dataset
has been processed using the standard STScI pipeline procedure. After
correction for geometrical distortion, images taken with the same
filter have been registered and combined to remove cosmic ray hits.
Total exposure times of the final combined images are $\sim$10,000~s
and $\sim$7200~s in F439W and F555W, respectively.  For each filter,
a set of short exposures was also obtained to provide photometric
measurements of stars which would saturate in the deep images.
Exposures of 20 s,  40 s, and 60 s duration were obtained for the
F555W, F439W, and F380W filters, respectively, at two different
pointings and are described in De~Marchi \etal\ (1997).

Many bright star clusters and several thousand stars are clearly
visible in the deep WF+PC images.  
The automated star detection routine DAOFIND was applied to
the data, by setting the detection threshold at $5\,\sigma$ above the
local background level. We carefully examined by eye each individual
object detected by DAOFIND and discarded saturated stars, a number of
features (PSF tendrils, noise spikes, etc.) that DAOFIND had
misinterpreted as stars, as well as a dozen extended objects (star
clusters) whose full width at half maximum (FWHM) exceeded that of 
typical stellar objects in PC frames (FWHM
$\simeq 1.6$\,pixel) by a factor of two or more.

In this way, a total of $\sim$7000 unsaturated stars were detected in
the field covered by the PC. The severe crowding of our images,
particularly towards the center of the galaxy, led us to prefer the use
of the PSF-fitting DAOPHOT photometry package over the simpler, but
more limited, aperture photometry techniques.  A sample PSF was built
by using three relatively isolated and moderately bright stars 
(\mb~$\simeq$~\mv~$\simeq$~21.5) located on opposite sides with respect to
the body of the galaxy. Instrumental magnitudes measured in this way
were converted into the WFPC2 synthetic system VEGAMAG as indicated in
Holtzman \etal\ (1995). Interested readers are referred to that paper  
for further details on the characteristics of the WFPC2 photometric
system. In this paper, \mb\ and \mv\ refer to the WFPC2
synthetic system. Although these magnitudes are very similar to Johnson
$B$ and $V$ magnitudes, no attempt has been made to convert our
magnitudes into the standard Johnson system.

The main source of uncertainty in our photometry is crowding,
particularly in the central regions of NGC\,1569, where the average
star-to-star distance is on the order of $\sim$4~pixels. The photometric
error estimated by DAOPHOT (\sigmada), and confirmed by artificial star
tests (see below), is smaller than $0.1$\,mag in both bands down to 
\mb~$\simeq$~\mv~$\simeq$~24, and it grows to $\sim$0.3~mag at magnitude
26 in both bands. The behavior of the photometric error with magnitude
is shown in Fig.~\ref{fig-sigma}.

We carried out a series of artificial star tests in order to estimate the 
fraction of stars lost to crowding as a function of their brightness. We
ran 10 trials in each of the two filters, by adding each time a number
of artificial stars not exceeding $10\,\%$ of the total number of
objects in the frame, to avoid overcrowding. The magnitudes of the
artificial stars were distributed according to the empirical luminosity
function.  Recursive DAOFIND runs, with the same parameters as in the
reduction of the scientific data, allowed us to assess the fraction of
added stars that were recovered and the magnitudes at which these
objects were detected. Since the artificial star routine (ADDSTAR)
places objects at random over the area of the frame, these tests
provide an average value of the photometric incompleteness in our
images. Artificial star tests were run on both filter frames, but the
F439W image, being less deep, in practice defines the overall
completeness limit. All artificial stars not recovered in the
F555W frames were also missing in the F439W ones. This is partly due to
the intrinsic color of the stars, and partly to the lower efficiency of
the instrumental setup (WFPC2 + F439W) in the bluer band.  Since the
assumed color distribution of the artificial stars follows the mean
locus observed in the color-magnitude diagram (CMD), the 
incompleteness factors
derived in the two filters are not independent and, as such, the
overall incompleteness fraction is set by the larger of the two values at
each magnitude.

For each artificially added star we compared the input magnitude with
that at which the star was recovered, and found no systematic
difference. In addition, the standard deviation of the recovered values
is in excellent agreement with the $1\,\sigma$ uncertainty of the
photometric reduction at any magnitude level.

Our simulations show that at \mb~$\simeq$~25.0 on average about 30\,\%
of the artificial stars added to the PC chip are not detected by the
automated routine, but this number goes down to only 8\,\% at 
\mb~$\simeq$~23.5. At magnitude levels brighter than \mv~$\simeq$~23 all
artificial stars are detected, except when they fall on top of
one of the bright super star clusters. Our photometry is, therefore,
$\sim$100\,\% complete for \mv~$<$~23.  Table~\ref{tab-compl} shows the
completeness factors as a function of the F439W magnitude, as
determined by the artificial star tests. These are the values adopted
in our theoretical simulations.
 
The artificial star experiments provide us with a statistical estimate
of the number of stars lost to crowding.  There is, however, another
issue that must be considered when comparing the observations with
theoretical simulations: blending affects
the derived magnitude of detected objects.  The extent to which the
images of two stars can merge with each other, thereby resembling a
single object with photometric properties different from those of the
parent stars, is a function of their angular separation and of their
magnitude difference. If the projected angular distance between two
stars is smaller than the PSF-FWHM, they will be detected as a single
object by the photometric routine, and the number of stars lost because
of this is included in the incompleteness factors. Concerning the
photometric error, we select objects with \sigmada $<$ 0.1\,mag and take 
0.1\,mag as an estimate for the photometric accuracy of the data.
Thus, we are essentially interested in those cases in which the
uncertainty introduced by blending on the determined magnitudes is
larger than this value. This occurs when the blended stars differ in
magnitude by less than $\sim$2~mag.

Our tests show, however, that even in this case the sharp PSF of the PC
camera considerably limits the effects of blending, in spite of the
high density of stellar objects in our frames. The results of
the artificial star experiments, as well as statistical considerations,
indicate that the probability of blending for a pair of stars differing
less than 2\,mag in flux amounts to $\lesssim 1\,\%$ for objects
brighter than \mv~$\simeq$~25. Therefore, it is unlikely that blending
introduces a large photometric uncertainty. Blending of objects with a
large magnitude difference is much more frequent. In this case, the
fainter component is virtually lost with a small effect on the brighter
one, and this is accounted for in the simulations by correcting for
photometric error and incompleteness.

Two of the WF frames have also been reduced, the third having been
discarded because of the presence of a highly saturated star.
We found $\sim$1000 objects
with \sigmada $<$ 0.2\,mag in both the F555W and F439W filters. 
Since the general characteristics of the stellar distributions in 
the CMDs derived for the WF and for the PC frames are very similar, we
proceeded with a quantitative analysis only for the CMD derived from
the PC frames. 

\section{Color-magnitude diagram and luminosity function}

Fig.~\ref{fig-cmdo} shows the CMD in the HST F439W and F555W 
bands resulting from the analysis of the region of NGC~1569 contained in the 
PC field of view. As discussed by De Marchi \etal\ (1997), this region
contains three super star clusters (SSCs), two of which form a close pair which
is barely resolved. Only integrated photometry is feasible for
the SSCs, which are not resolved into stars. The resulting $(B-V)$ and
$(U-B)$ colors are indicative of young ages, but is not possible
to date the clusters precisely. 

The diagram in panel a) shows the 2809 objects measured in both 
the F555W and F439W bands with a photometric error \sigmada~$<$~0.2,
while the diagram in panel b) shows the subsample of 801 objects with
\sigmada~$<$~0.1. 
By comparing the two panels it is apparent that the major features of the
galaxy stellar population are well delineated in both cases. 
Clearly, the diagram with the
tighter selection criterion contains a lower number
of stars because it reaches a brighter magnitude limit, and because
it rejects most of the objects with extreme colors.

In the CMD shown in panel a) there are some very blue objects,
populating the bottom left corner.
Taking into account the average extinction affecting our field,
objects appearing with 
(\mb~--~\mv)~$\lsim$~--0.5  have intrinsic colors bluer than $\simeq$~--1. 
However,
neither galactic nor extragalactic objects can have colors that blue:
both white dwarfs with magnitudes in the relevant range, and the
bluest QSOs have intrinsic $(B-V) \gsim -0.6$.  
We have thus examined in detail all the 72 objects with 
(\mbv)~$<-0.5$  and photometric error \sigmada $<$ 0.2\,mag in both  
bands. They are all concentrated in the region of the PC field
dominated by the main body of NGC~1569, suggesting that
these extremely blue objects originate from phenomena
associated with the galaxy (either physical or spurious), and not from
background or foreground effects. A detailed examination in all of the
three photometric bands revealed that most of these objects are
spurious detections, 
due to noise peaks falling on the wings of stellar images, 
or to the blend of 2 stars. When selecting only the objects with
errors \sigmada $<$ 0.1\,mag no point source with extremely blue
color is found.
Since the shape and position of the bulk of the blue plume and of the red tail 
look unaffected by the selection criterion, we will adopt 
the diagram in panel b) as reference observational CMD for the 
stellar population in the field of NGC~1569.

The general morphology of the CMD of NGC~1569 is similar to that of 
Local Group irregulars observed from the ground (see, e.g., Freedman 1988; 
Tosi \etal\ 1991, hereinafter TGMF; Greggio \etal\ 1993, hereinafter
GMTF; Marconi \etal\ 1995, hereafter MTGF; Gallart, Aparicio, $\&$
V\'{\i}lchez 1996b), 
with a quite  
scattered distribution and a prominent concentration of stars in the 
blue plume. The CMD shows a smaller fraction of red stars, but this is
mostly due to the selection effect of the magnitude limit for object
detection in the F439W frame. Indeed, the $I$ vs. $(V-I)$
diagram of NGC~1569 obtained VB shows
a significant population of red (mostly intermediate age) stars. 

The median color of the blue plume (\mbv)~$\simeq$~0.45 is quite red as a 
consequence of high foreground extinction. By correcting this 
color for the average foreground reddening $E(B-V)=0.56\pm0.1$ as 
estimated from UV observations by Israel (1988a), an intrinsic 
(\mbv)$_0$~=~--0.1 is derived. This is equal to 
or slightly redder than the median $(B-V)_0$ color 
of the blue plumes in other nearby irregulars (see, e.g., MTGF and Gallart 
\etal\ 1996b for NGC 6822; TGMF for Sextans B; Bresolin 
\etal\ 1993 and GMTF for NGC 3109). 
However, since Gonz\'alez-Delgado \etal\ (1997) 
have derived $E(B-V)=0.67\pm0.02$
for the super star clusters in NGC~1569 (using UV, 
optical and near-IR spectra), it may well be that this galaxy is affected
by internal differential reddening. Thus, the intrinsic colors of the
blue plume stars could be bluer than the previous estimate.

Given the low galactic latitude $b$~=~+11$^\circ$ of NGC~1569, 
the observed CMD 
is likely to be populated by a significant number of foreground stars. 
Despite our original 
request, the time allocated on HST was not sufficient to take exposures
of adjacent fields and measure directly the Galaxy contamination. 
Since the CMD derived from the WF images looks very
similar to that derived from the PC frames, the galaxy extends substantially
over the WF frames, which cannot then be safely used to discriminate
between members and foreground/background objects. 
However, using ground based observations, VB derive a very low density
of external field objects in the direction of NGC~1569.
Thus our CMD
should be almost unaffected by foreground/background contamination.

The full dots in Fig.~\ref{fig-fl} represent the luminosity function (LF)
of the 801 selected stars of NGC~1569, whereas the open circles correspond
to the stars with 0~$\leq$~(\mbv)~$\leq$~0.5, which in the following will be
referred to as the
{\it reference blue plume}. This subsample has been selected
simply for the purpose of comparing our results with those relative to other
galaxies and/or other authors, and does not represent
either the whole blue plume or the pure main sequence (MS) stars.

In the past, the LF of the blue
plume was usually adopted for estimating the mass function, under
the assumption that it is essentially populated by MS stars.
However, as already noticed for other irregulars (e.g., Freedman 1988; 
TGMF and GMTF; Gallart \etal\ 1996b), the comparison of the observational 
CMD with theoretical stellar models demonstrates that the 
blue plume is populated also by a significant
fraction of stars at the hot edge of the blue loop evolutionary phase. 
This represents a serious problem for authors deriving the initial
mass function (IMF) from the 
luminosity functions of the blue plume assuming it contained only MS stars. 
The MS--post-MS blend is apparent also in the $I$ vs. $(V-I)$ 
diagrams 
described by VB. As discussed there and in the next 
section, it is virtually impossible to discriminate MS from post-MS stars at 
the brightest magnitudes, where the morphology of the theoretical 
evolutionary tracks of massive stars in the CMD leads to the merging
of the blue
edges of the loops into the region populated by MS stars. 
Our (\mbv) selected {\it reference blue plume} is therefore just a 
compromise between the 
requirement of removing as much as possible the evolved stars and the 
need of keeping the brightest MS objects.

The slopes of the LFs are usually derived with least squares linear fitting to
their complete portions. However, as shown in GMTF and MTGF,
the large statistical fluctuations due to the small number of sampled
stars, especially at the brightest magnitudes, may significantly alter
the results of a method requiring data binning. To overcome this problem, 
we have applied a maximum likelihood fitting to evaluate 
the slope of the LF in the {\it reference blue plume}.
Down to \mv~=~23, where our sample is complete (see previous section),
the slope turns out to be $\Delta$ log $N/ \Delta V = 0.66 \pm 0.05$. This
value is higher than $0.46\pm0.06$ derived by VB,
a result of the different methods: for instance,
a linear fit gives a flatter slope of $0.6\pm0.1$ for our data set.
Furthermore, the fit of VB is not restricted to 
the data portion where completeness is reached.

Our slope $\Delta \log N/\Delta V = 0.66 \pm 0.05$ is
consistent with the average $0.70\pm0.03$ derived by Freedman (1985) and
Hoessel (1986) from a large sample of irregulars.  
It appears, however, slightly steeper than the slopes
derived by us in a homogeneous way (see TGMF; GMTF; MTGF) for
four nearby dwarf irregulars, whose average value is 
$\Delta \log N/ \Delta V \simeq 0.55$.

\section{Theoretical models}

The observed CMD and the (total) LF described in the previous section have been
compared to corresponding theoretical simulations, based on stellar
evolutionary tracks. This method has proven to be a powerful tool
to investigate the star formation history of galaxies with young
stellar populations (see Tosi 1994; Greggio 1994). 

Synthetic CMDs are constructed via Monte~Carlo extractions of 
(mass, age) pairs, according to a prescribed IMF and a star-formation rate
(SFR) law.
The extracted object is placed on the theoretical H--R diagram (HRD)
by interpolating on
evolutionary tracks, and its luminosity and effective temperature are
converted into magnitudes using tables for bolometric corrections and
colors. A photometric error and incompleteness test, both as measured on
the observational frames, is finally applied, and the synthetic star
is placed on the CMD. 
Once the number of objects populating the synthetic CMD equals
that of the observed one, the procedure is stopped, yielding the
quantitative level of the SFR consistent with the observational data,
for the prescribed IMF and shape of the SF law. 
This procedure automatically accounts
for the stochastic nature of the SF process, small
number statistics, photometric errors, and incompleteness factors
affecting the real data. In addition, it easily allows us to explore
the effects of using different sets of stellar tracks.

We have considered several sets of stellar tracks (Table~\ref{tab-mod})
 in order to compare their effect
on the derived SF history, and to check their adequacy in reproducing
the main features of the observed CMD.
The interpolation algorithm was carefully chosen
to reproduce the main characteristics of the
evolutionary sequences. Due to numerical effects, the short lived, post-MS 
phases may be underrepresented in the Monte~Carlo
simulations. Thus, the tracks have been trimmed in subphases, and for
each extracted object the evolutionary subphase is first determined.
Interpolation between adjacent tracks is then performed within the 
appropriate subphase.
This procedure ensures that the main features of the synthetic
diagrams, (e.g., the blue-to-red
ratio, the ratio between the MS and post-MS objects, the
number of W-R stars, etc.) are consistent
with the adopted set of tracks. 

The luminosity and effective temperature of the synthetic objects are
converted into magnitudes and colors by interpolating on 
tables constructed from the collection of model atmospheres by
Lejeune, Cuisinier, $\&$ Buser (1997),
which cover (although non-uniformly) the effective temperature and
gravity ranges: 2000~K~$<$~\TEFF~$<$~50,000~K,
--1.02~$<$~log~$g$~$<$~5.5, and are characterized by several 
values of the metallicity ([Fe/H]), with solar elemental ratios.
It should be noticed here that the tracks used for the simulations
do not reach very low gravities, or temperatures lower than $\sim$3600~K. 
Therefore, the model atmospheres used for our
synthetic diagrams are essentially
those of Kurucz (1992). The sets at metallicities [Fe/H]~=~--0.5 and
[Fe/H]~=~--1 have been selected for our application. 

The conversion tables have been prepared as follows. For each model
atmosphere in the grid, a theoretical bolometric correction
\BCvth~ has been computed: 
\begin{equation}
\BCvth = -2.5 \log \frac {\int_{V}
{L_\lambda}t_{\lambda}d\lambda} {L_{bol}}  \label{eq:bcv}
\end{equation}                   
where $t_\lambda$ is the $V$-filter transmission function in the VEGAMAG
system. Similar tables for the \mv$-V$ and \mb$-V$ colors
(in the VEGAMAG system) have been
produced by processing the model atmospheres through the CALCPHOT
routine in the SYNPHOT package. 
For the objects populating our synthetic HRD, the following relation applies
\begin{equation}
M_V = \BCvth - 2.5 \log L_{bol} + C_V \label{eq:mag}
\end{equation}
where $C_V$ is a calibration constant. In order to ensure that the
theoretical magnitudes are in the {\it same photometric
system as the observational CMD}, $C_V$ has been
fixed by enforcing that for the solar model
atmosphere (\TEFF~=~5777~K,
log~$g$=~4.4, [Fe/H]~=~0.), we obtain $M_V$~=~4.7785
if we adopt  $L_{bol,\odot}=3.96\times10^{33}$~erg~s$^{-1}$. 
This is the value  
in the VEGAMAG system found by SYNPHOT when {\it observing} a star with
the same flux distribution, the solar radius, at a distance of 10 pc.
In this system, the solar model and the
model atmosphere representing Vega (\TEFF~=~9400~K, log~$g$~=~3.9, 
[Fe/H]~=~--0.5) have (\mb~--~\mv)~=~0.717 and 0.0066, respectively.

For each synthetic object, characterized by (mass, log~$L$, log~\TEFF), the 
theoretical bolometric correction and colors are determined through
interpolation in these tables. $M_V$ is then derived from 
eq. (\ref{eq:mag}), and henceforth
also the HST magnitudes in the VEGAMAG system.

\section{Synthetic CM diagrams}

The synthetic CM diagrams (CMD) and LFs depend on the input parameters
in the simulation: IMF, history of the SFR, distance, reddening and
theoretical stellar evolutionary tracks. Adopting a value for the
distance modulus and reddening,
the superposition of the evolutionary tracks to the observed CMD
yields the range of ages and masses sampled by our data. As an example,
in Fig.~\ref{fig-cmdth001} we show the Geneva tracks with Z=0.001
trasformed to the observational plane, having adopted the 
canonical values $(m-M)_0=26.71$, $E(B-V)=0.56$ from the literature.
Comparing Fig.~\ref{fig-cmdth001} with Fig.~\ref{fig-cmdo}, panel a),
one derives that
the bluest stars in the blue plume are MS
objects with masses larger than $\sim$ 9 \MSUN, and then ages younger
than $\sim$ 30 Myr. The wide color distribution 
shows the presence of evolved stars, with masses larger than $\sim$
4 \MSUN, and then ages younger than $\sim$ 0.15 Gyr. Thus, SF has 
been active in our field over $\sim$ 0.15 Gyr. Although the smoothness of the
star distribution in the CMD qualitatively suggest a continuous kind
of SF, 
this issue can only be quantitatively assessed computing the simulations,
which take into account the smearing effect of photometric errors and
the incompleteness at faint magnitudes. In our procedure,
then, we first try the simplest possible case, i.e. a constant SF over
the look-back time, and then vary the SFR shape and IMF slope to reproduce
the data distributions. When an acceptable representation of the
observation is reached, we perturb the parameters to test the
stability of the solution.

The comparison of the theoretical simulations with the observations is
performed quantitatively on the LF, and mostly qualitatively on the
appearance of the CMD. A detailed quantitative comparison of
the color distribution of stars (like e.g., Tolstoy $\&$ Saha 1996) 
would not add much
information, due to the smearing effect of photometric errors, and, more
importantly, to the intrinsic uncertainty affecting the effective
temperatures of models of massive stars in the post-MS
phases (Renzini \etal\ 1992). Small variations of the theoretical
\TEFF\ 
result in large differences of the predicted distribution of objects
throughout the CMD, due to the high sensitivity of colors and
bolometric corrections to the effective temperature in the relevant range.

Because of systematic uncertainties, there is often 
no significant difference between the best simulation and other similar
solutions. Rather, our aim is to isolate the range of IMF and SFR
parameters which are consistent with the data.
We have computed a few hundred synthetic CMDs varying the input
parameters mentioned above.
In the next subsections we discuss our results and the influence of
the input parameters on the conclusions. 
We start with adopting a distance of 2.2 Mpc, and present our preferred 
solutions for various sets of evolutionary models. Second we discuss 
how different SF histories influence the results. Finally we consider 
the effect of an uncertainty in the distance.
The reddening is varied, within the observational estimate, in such a
way that the average color of the synthetic blue plume agrees with the 
observations.

\subsection{Influence of the Evolutionary Sequences.}

The SF history derived from the analysis of an observational CMD
depends on the set of tracks adopted in the simulation (Greggio 1994).
We have therefore considered several sets of stellar tracks
(Table~\ref{tab-mod}). Although the metallicity of NGC 1569 is
Z $\sim$ 0.004 we have computed simulations based on the
Geneva tracks with several values of $Z$.
This is motivated by the fact that
the overall appearence
of the CMD is sensitive to the morphology of the blue loops, which
in turn
critically depends on Z (Ritossa 1996).

We have also explored the difference in the derived interpretation
when adopting the Padova set of tracks at Z = 0.004.
The main difference between the Padova and the Geneva sequences concerns
the treatment of convective overshooting: both sets include core
overshoot, but to a lesser extent in the Geneva models. In addition, the
Padova tracks also include overshooting from the base of the
convective envelope.  

It is worth recalling some general features of
the evolutionary sequences: \par\noindent 
i) All sets are characterized by extended loops
for masses above 4~\Ms\, with the maximum extension for the 
Geneva $Z$~=~0.001 and Padova $Z$~=~0.004 sets. \par\noindent
ii) Among the Geneva tracks, the fraction of evolutionary lifetime
spent in the blue portion of the CMD   
during the post-MS phase
is largest for the $Z$~=~0.001 tracks (actually close to 100$\%$ for stars 
more massive than $\sim$4~\Ms), and it decreases with increasing 
metallicity.
The Padova set at $Z$~=~0.004 is characterized by a value of this ratio
which is intermediate between the Geneva $Z$~=~0.001 and $Z$~=~0.004 sets.
\par\noindent
iii) The ratio between post-MS and MS evolutionary lifetimes for stars
more massive than $\sim$8~\Ms\ is $\sim$0.1 for the Geneva tracks. The
same ratio is
considerably smaller for the Padova tracks, reaching a minimum value of
$\sim$0.06 for the 12~\Ms\ track.

These characteristics crucially determine the main differences in the
simulations constructed with the various sets of tracks.

\subsubsection {Simulations with the Geneva tracks with $Z$~=~0.008,0.004}

Adopting the Geneva tracks with $Z$~=~0.008 and $Z$~=~0.004
we were not able to reach a satisfactory agreement between the
synthetic diagrams and the observations, regardless of the choice of the IMF
and SF parameters. As an example, in 
Fig.~\ref{fig-synth1fl} we show two of the best cases obtained with
$Z$~=~0.008 (panel a) and $Z$~=~0.004 (panel b). The corresponding LFs are
shown in panel c), with the solid line referring to the simulation displayed
in panel a). The adopted reddening values are $E(B-V) = 0.46$
and 0.56 for panels a) and b) respectively, which compensate for the 
different metallicities of the tracks, and are both consistent with
Israel's (1988a) estimate.   

The CMD in panel a) has been computed with two partially overlapping 
SF episodes,
so that the SFR peaks around $\sim$0.05~Gyr ago. The adopted IMF
exponent is $\alpha=3$ (in these units the Salpeter's exponent is 2.35). 
The simulation in panel b) adopts a mildly decreasing SFR, and
a Salpeter IMF. In both simulations the synthetic blue plume extends 
up to too bright magnitudes and the bright portion of the LF is
overpopulated over a wide magnitude range. These characteristics
appear in all the simulations with this sets because they
are related to the shape of the tracks on the CMD. Infact,
due to the relatively short color extension of the blue loops of
models with mass around 5 \MSUN (see Fig.~\ref{fig-cmdth008} for the
$Z=0.008$ set),  
the faint portion of the blue plume has to be mainly
populated by MS stars. This is a consequence of the photometric 
incompleteness in
the F439W frames, which depletes the number
of evolved stars when their colors are too red.
Thus, with these tracks a large number of massive
stars is required, whose 
post-MS progeny shows up as bright evolved supergiants. 
In the specific case of the $Z$~=~0.008 tracks, 
at magnitudes brighter than \mv~$\simeq$~21.5 we find post-MS objects 
with masses between 15
and 25~\Ms , which are the evolved progeny of fainter MS objects populating the
blue plume in the range \mv~$\simeq$~25~--~22.5. 
Thus, the excess of bright objects cannot be suppressed without 
substantially underpopulating the faint portion of the blue plume, and
could only be avoided by invoking such a delicate fine-tuning of the SF
behavior, to appear very contrived. 
Similar arguments apply to the $Z$~=0.004 set, in which the extension
of the blue loops of the 5 \Ms\ and 7 \Ms\ tracks is so short, that
the Herztsprung gap appears at the faint magnitudes, in spite of the
large photometric error.

The CMDs in Fig.~\ref{fig-synth1fl} were computed with the low mass
loss option in the
Geneva models. Switching to the high mass-loss sets
does not improve the comparison between the simulations and the
data. For $Z$~=~0.008, the higher mass loss only affects 
the location of the core
helium burning phase for the more massive models. As a consequence,
the excess objects are still present, yet shifted to intermediate and red
colors. For $Z$~=~0.004, the adoption of a high mass loss rate would only
affect the evolution of the most massive stars ($M \gsim 40$~\Ms\ ),
spanning
a mass range which is almost not populated in our CMD.  

\subsubsection {Simulations with the Geneva tracks with $Z$~=~0.001}

As can be seen in Fig.~\ref{fig-cmdth001}, the
loops for the Geneva tracks with $Z$~=~0.001 extend to very blue colors, 
and a substantial fraction of
the faint blue plume can be populated by relatively low mass
stars in their core helium burning phase. 
Unlike the previous cases, we find satisfactory 
agreement with the observed CMD
and LF with various combinations of the IMF slope and SFR, provided
that the activity ceased $\sim$7~Myr ago (corresponding to the
evolutionary lifetime of a $\simeq$ 30 \Ms\ star) to avoid an excess of
bright stars. Fig.~\ref{fig-synth2fl} shows our best models
for a constant SFR
(panel a) and an exponentially declining SFR with an e-folding time of
0.6~Gyr (panel b). Notice that this latter case implies a variation of
the SFR of only $\sim$ 20$\%$ over our short look-back time.
The corresponding LFs are displayed in panel c). A reddening of
E(B-V)=0.56 has been adopted for these simulations. It can
be seen that both cases give a very good fit to the observed LF.
We  adopted a steeper IMF for the constant SF case, in order to
avoid too many bright stars. The synthetic CMDs also give a fair
representation of the data, except for a slight paucity of stars at
intermediate and red colors. 

In order to see whether the data allow for large variations 
of the SF activity during the last 0.15 Gyr we have computed
simulations with two separate SF episodes. Fig.~\ref{fig-synth3fl} shows
our best models obtained by assuming that the SFR in the most recent
episode is higher (panel a), and lower (panel b) with respect to the
previous episode. In both cases the SFRs in the two episodes differ by
a a factor of $\sim$ 1.2.  
Again, we achieve a fair representation of the data when appropriately
tuning the IMF slope with the SF history. 
In all acceptable simulations, however, the SFR in the
last 0.15~Gyr has proceeded approximately at the same level, and the
slope of the IMF is close to Salpeter's value or slightly
steeper. 

Our procedure corresponds to counting and weighing the stars which
appear in the observational CMD, and determine their ages. 
The derived level of the SFR is then relatively insensitive to the
combination of IMF and SF history adopted in the
simulations. Indeed, our models in Fig.~\ref{fig-synth2fl} and
Fig.~\ref{fig-synth3fl} show agreement with the
observations, in spite of the different prescriptions. Yet, the
average values of the SFRs over the last 0.15 Gyr, in stars more
massive than 4 \MSUN, are 0.07 and 0.09 \Myr for the simulations in
Fig.~\ref{fig-synth2fl}
(panels a and b respectively), and 0.08 and 0.11 for the simulations in
Fig.~\ref{fig-synth3fl} (panels a and b respectively). This applies
only to
the mass range directly observed. When extrapolating the SFR for
stars with masses ranging from 0.1 to 120~\Ms , the average rates
become $\sim$ 3 \Myr for $\alpha=3$ and
$\sim$0.5~\Myr, for $\alpha=2.35$.

\subsubsection {Simulations with the Padova tracks with $Z$~=~0.004}

Unlike the Geneva tracks with the same metallicity, the Padova tracks
with $Z$~=~0.004 are
characterized by well extended loops for masses larger than $\sim$5~\Ms\, 
whose blue edges almost merge into the MS. As a consequence, the
faint blue plume is populated by MS and helium burning stars.  
Moreover, in the range 9~--~25~\Ms\ 
the ratio between post-MS and MS lifetime is significantly smaller than in
the Geneva tracks, implying a lower ratio between bright
evolved supergiants and their MS progenitors.
For these reasons the synthetic diagrams with this set
of tracks are fairly different from those obtained with the Geneva
tracks with the same metallicity.
We are able to obtain a fair representation  
of the data,
provided that the activity stopped $\sim 8-10$~Myr ago to avoid an excess
of bright stars. By adopting the spectroscopic value for the reddening,
$E(B-V)$=0.56,
the color of the synthetic blue plume appears somewhat redder than
the observed one. We have therefore computed cases with both 
$E(B-V)=0.56$ and 0.46. 

Although assuming a constant SFR we obtain acceptable simulations, the
best fitting models are characterized by two distinct episodes of SF,
the most recent one starting $\sim$ 30 Myr ago.
We find
acceptable solutions with SFRs both increasing and decreasing with
time, in combination with different values of the IMF slope and
adopted reddening. However, all the solutions indicate that the SFR has
not varied by a large amount within the last 0.15~Gyr, and that the
IMF slopes are slightly steeper than Salpeter's value, and
close to that derived for solar neighborhood stars (e.g., Scalo 1986).
As an example, Fig.~\ref{fig-synth4fl} shows the synthetic CMDs
obtained with two bursts, the most recent one occurring at a lower rate in
both panels a) and b).
The theoretical and observed CMDs and LFs are in good agreement,
perhaps with a slight excess of red supergiants in the models.  
In these models the ratio between the SFR in the older and most recent
episode is 1.5 (panel a) and 1.2 (panel b). Thus, also with this set 
of tracks the data are consistent with an almost constant SFR over the 
last 0.15~Gyr (within
20$\%$ to 50$\%$). The average level of the SFR in the simulations 
in Fig.~\ref{fig-synth4fl} are 0.1, 0.07 \Myr in panels a and b
respectively, in stars younger than 0.15 Gyr and more massive than 4
\Ms . When extrapolating the IMF down to a lower mass cut-off of 0.1
\Ms\ we obtain an average SFR of is $\sim$1~\Myr if $\alpha=2.6$, or 
$\sim$3~\Myr\ if $\alpha=3$. This holds 
when adopting $E(B-V)=0.46$. For the higher 
reddening the derived rates are $\sim$ 1.3 times larger.

\subsection{Testing the SF history.}

Given the smooth distribution of stars in the CMD, and the shape of
the LF, large variations of
the average SFR within the last $\sim$ 0.15 Gyr are not consistent
with the data, as we have tested numerically. 
In this section we
rather concentrate on the epoch of the recent stop in the SF, and on
the possibility of short quiescence periods.   

Fig.~\ref{fig-synth10fl} shows the sensitivity of the results on the
adopted age for the recent stop in the SF in the case of the Geneva
tracks with $Z$~=~0.001. In the bottom panel the SF
has ceased 5 Myr ago, while in the top panel it is still
ongoing. A steep IMF has been adopted for these simulations, in order
to minimize the number of massive stars. Despite that, an
ongoing SF predicts an excess of bright objects, while a 5 Myr stop is
consistent with the data.
Fig.~\ref{fig-synth6fl} panel b) shows the case of a constant and 
currently ongoing SF
having adopted the Padova $Z$~=~0.004 models.
In spite of the steep exponent of
the IMF, the CMD turns out to be overpopulated at the bright end,
over a range of $\sim$1.5 mag. With this set of tracks, adopting
$\alpha$=3, we still find an excess of bright stars stopping the SF 7
Myr ago. It is worth noticing that the age of the stop is constrained
by the bright portion of the LF, over a range of $\sim$ 1 mag or more.

We now turn to consider the possibility of discrete episodes of SF,
separated by quiescent phases. In general, this behavior
has a definite signature on the CMD, i.e. the lack of objects around
the isochrones with ages corresponding to the quiescence period, which
shows up in the evolved portion of the diagram as an 
almost horizontal gap.
Due to the effect of photometric errors, however, this signature
can go undetected on the observed CMD. 

A quiescent period lasting from 35 to 30~Myr ago 
has been assumed in the simulations shown in 
Fig.~\ref{fig-synth4fl}, in order to improve the
appearance of the synthetic CMDs. The simulations with a
constant SFR are characterized by a clump of blue
supergiants at \mv~$\simeq$~22 (see Fig.~\ref{fig-synth6fl}, panel b) 
populated by objects with masses in the
range $\sim$9 to 12~\Ms. 
The blue loops of these sequences happen to converge to the same
location in the CMD, providing an excess of objects in this region. 
Isochrones with ages of 35 to 30~Myr are
populated by post-MS objects with masses in the range $\sim$ 9 to
10~\Ms\, and the
assumed quiescent period makes this clump of blue SGs less prominent.
It should be noticed that a slightly different shape of the loops of these
sequences would not produce this feature, making the quiescent period
unnecessary. We also find that
quiescent periods longer than $\sim$10~Myr should be excluded, since
they show up in the synthetic CMD with a very evident horizontal gap,
which is absolutely absent in the observed diagram. This is
illustrated in the top panel of Fig.~\ref{fig-synth6fl}, where we show
a simulation in which no SF has occurred between 40 and 30 Myr ago. 

The simulations in Fig.~\ref{fig-synth3fl}, based on the Geneva tracks
with $Z$~=~0.001,
have been computed assuming two contiguous SF episodes. Also for these
tracks we have tested the possibility of short interburst
periods. Again, we find
that quiescent periods longer than $\sim$10 Myr in the 
recent past (less than 0.1~Gyr ago) would be evident in the
observational CMD, and are then excluded by the smoothness of
the data.

\subsection{Dependence on the adopted distance.}

The distance to NGC~1569 is relatively uncertain. The most widely quoted values
in the literature are between 2.2 and 2.5~Mpc (Arp \& Sandage 1985; Israel 
1988a; O'Connell \etal 1994). They are based on the properties of the
still resolved brightest
individual supergiants. However, distances of up to
4~Mpc are possible since NGC~1569 is a probable member of the IC~341/Maffei~1
\& 2 group, whose distance is uncertain. If NGC~1569 were at
4~Mpc, it would be quite extraordinary in several respects. Apart from
overluminous individual supergiants it would harbor the most luminous resolved
super star clusters known in the universe. The brightest cluster has
$M_V = - 13$ to $-14$ already for the shorter distance (O'Connell \etal 1994;
De~Marchi \etal 1997). At 4~Mpc, this cluster would have appeared at 
$M_V \simeq -17$ if observed at peak brightness 8~Myr ago! A larger distance
would make the properties of the ring nebula in the outskirts quite
unusual as well. Drissen \& Roy (1994) noted that its size and \Halpha
luminosity
exceeds those of similar ring nebulae in M33 by up to an order of magnitude.
This apparent discrepancy would be even more amplified for the larger distance.

While none of these arguments by itself provides convincing evidence for or
against one or the other distance, taken together they all point toward the
shorter distance. 
However, a larger distance would imply younger ages for the
stars in our observed CMD, as well as larger levels of the
recent SFR. To test the effect of the uncertainty in the distance to
NGC~1569 we have computed simulations assuming a distance of 4
Mpc. The observed CMD now samples stars younger than $\sim$ 50 Myr.
Two of our best cases are shown in Fig.~\ref{fig-synth8fl}, 
for the Padova
tracks with $Z=0.004$. The simulations assume a still ongoing SF (bottom
panel) and a recent stop, at 5 Myr (top panel), and again the slope
of the IMF has been chosen to minimize the number of massive stars.
It can be noticed that, even adopting a large distance, a stop in the
recent SF activity is preferred, not to overpopulate the bright end of
the LF. The level of the SFR in these simulations is higher by a
factor of $\sim$ 2.8 with respect to what derived with the shorter
distance. 

\section{Discussion}

\subsection{Summary of the main results}

In the previous section we have shown that
the Geneva tracks with metallicities $Z$~=~0.008 and 0.004 do not allow us to
reproduce the main features of the observed CMD and LF, regardless of
the adopted combination of IMF and SFR. On the contrary, 
we find a fair agreement
between the simulations and the data when adopting either the Geneva tracks
with $Z$~=~0.001, or the Padova tracks with $Z$~=~0.004, the latter value being
consistent with the metallicity derived observationally.
These results, however, are mostly based on the shape of
the blue loops, which are intrinsically uncertain.
Therefore, we can neither conclude that there is an 
indication for a
lower metallicity in this galaxy, nor that the input parameters used
in the Padova tracks are to be preferred to those in the Geneva tracks, as 
our analysis does not allow us to constrain the shape of the tracks
down to this
level of detail. Rather, we maintain that both sets of tracks have the
combination of input physics and metallicity parameter that yields a
good representation of our data. 
The color distribution of the evolved stars cannot be used to
determine the metallicity of the system, other than to derive 
a general ranking.
 
We can derive several consistent indications on the SF history in
the field of NGC~1569, regardless the adopted sequences: \par\noindent
i) The SF activity stopped 5--10~Myr ago.
\par\noindent
ii) During the last 0.1--0.15~Gyr the SFR has been approximately
constant. If quiescent periods have occurred, they did not
last longer than $\sim$10~Myr. \par\noindent
iii) The best agreement is generally reached with IMF slopes steeper
than Salpeter's , but the latter would not be inconsistent with the
data.
\par\noindent
iv) The average SFR in our field 
is approximately 3, 1, and 0.5~\Myr\ for IMF slopes of 3, 2.6,
and 2.35, respectively. These values refer to stars with masses ranging
from 0.1 to 120 \Ms\, but it should be emphasized that our data only
constrain the star formation of objects more massive than $\sim$4.5~\Ms. 
The dependence of these derived values on the adopted set of
tracks, reddening and shape of the SF law is found to be small. 
\par\noindent
v) Assuming that the minimum mass for a type II SN explosion is 8
\Ms\, the derived SFR implies that in the last 0.1~Gyr in our
region about $2-3 \times 10^5$ type II events have occurred, for an IMF
slope ranging from 3 to 2.35.

It may seem that enhancing the recent SFR while steepening the IMF
would lead to the same result, and to some extent this is true: enhancing
the recent SF would populate the CMD with young, and therefore massive,
stars, thereby counterbalancing the effect of a steepening of the
IMF (see Fig.~\ref{fig-synth2fl} and Fig.~\ref{fig-synth3fl}). However,
we find that acceptable solutions have parameters in a
relatively small range. In other words, when the shape of 
the SFR is radically different from the one giving a fair solution,
it is difficult to obtain acceptable simulations by varying the IMF slope
(and vice versa). Indeed, the two parameters do not act in precisely
the same way, as strengthening the recent SF with respect to
the past enhances the population along the younger isochrones by a
constant factor. Conversely, varying the slope of the IMF affects the
distribution of stars along all the isochrones.
Thus, for example, when the IMF slope is too flat, the LF of the
synthetic population is so skewed towards the bright end that only
very complicated SF histories may fit the data.  
As can be seen
from Figs.~\ref{fig-cmdth008} and \ref{fig-cmdth001}, the shape of the 
tracks, and therefore of the isochrones as well, is
such that in a given magnitude bin we find stars of different
generations and different masses, so that it is not possible to
disentangle the effects of SFR and IMF.  
The number of stars falling within suitably chosen boxes in the CMD 
(e.g., Bertelli \etal\ 1992) may  
allow this task. However, it is very hard to define these suitable boxes
in this case, due to the shape of the 
tracks for massive models in the observed CMD and to the 
photometric errors. 
This cautionary note applies in particular to the splitting of the
population of the blue plume into the MS and post-MS
contributions. 
For these reasons we need theoretical simulations, rather than
isochrone superposition, to interpret the data.
Moreover, as shown by our experiments, the 
shape of the blue loops heavily affects the synthetic CMD and LF so
that the final solution depends on the adopted set of tracks.  
The flexibility of the simulation code allows us to easily explore
this parameter space as well.

It is also worth noting that the SF stop that we infer
in the field of NGC~1569 is not particularly sensitive to the adopted
upper mass cut off for the IMF. For example, when the mass distribution is
truncated at $\sim$40~\Ms\, the synthetic diagrams obtained with an
ongoing SF are still overpopulated at the brightest
magnitudes. Indeed, lowering the mass cut off does not appreciably 
alter neither the slope of the LF, nor the overall ratio between the objects
populating the fainter and brighter portions of the LF. Conversely, 
when stopping the SF activity a few Myr ago, the whole MS population
is reduced with respect to the evolved one. 

Our quoted values for the SFR refer to a single-slope IMF
with stars distributed from 0.1 to 120~\Ms, and their dependence 
on the lower mass cut-off is of course very high. 
For example, the mass in stars
within 0.1 and 1~\Ms\ is $\sim$0.55, 0.75, and 0.9 of the total for
$\alpha=2.35$, 2.6, and 3, respectively.
Furthermore, these values depend on the IMF slope at the low mass end.
In a recent study by Gould, Bahcall, $\&$ Flynn (1997), the IMF for 
M dwarfs in the galactic disk can be fitted by a two-slopes IMF, with
a break at $M\simeq 0.6$~\Ms\, and with a slope $\alpha = -0.56$ (in
our usual units) in the low mass range. When adopting these
prescriptions for the low mass end of the IMF, and varying the slope
of the mass distribution only over the range $M\ge0.6$ \Ms\ , our 
average SFRs  become $\sim$0.8, 0.5, and 0.3~\Myr\ for slopes of 3, 2.6, and
2.35, respectively.

\subsection{Comparison with previous work}

The only previous study primarily dedicated to the resolved stellar 
population in the field of NGC~1569 is that by VB. 
By using pre-refurbishment WFPC images, these authors derive the CMD in 
the F555W and F785LP filters, roughly equivalent to the $V$ and $I$ bands.
The method used for the interpretation of the data relies, like ours,
on simulations of CMDs based on evolutionary tracks, but without including the
effect of photometric errors and incompleteness.
An additional difference concerns the use of the LF as
a tool to discriminate between possible models. 

VB conclude that 
NGC~1569 experienced a global burst of SF from 0.1~Gyr to 4~Myr ago,
and that a previous episode of SF had taken place in the galaxy, at a
substantially lower rate. Our observations are not deep enough in the
F439W band to detect the AGB stars
that reveal this older episode in the $I$ vs $(V-I)$ diagram.
As for the recent SF history, our results are roughly consistent with
their finding. We prefer an earlier stop of the SF activity, since,
when adopting an SF rate ongoing up to 4 Myr ago, the synthetic LF
tends to be overpopulated at the bright end, especially for the Padova
tracks with Z=0.004, which is the same set adopted by VB.
VB could not 
establish quantitatively this effect without the comparison of the
theoretical and observed LFs. In addition, we believe that
the photometric incompleteness prevents the determination of the epoch at
which the burst started, since stars older than 0.15$-$0.1 Gyr would
go mostly or totally undetected in our images. 
Finally, the level of the SFR of the recent burst that we derive using
the same set of tracks and IMF is consistent with theirs, judging from
the estimate of the recent SNII activity. 

To summarize, our results are in general agreement with
VB, but we believe that our conclusions are more
robust, due to both the superior quality of the refurbished HST data,
and to the method employed for the interpretation. 
 
\subsection{NGC~1569 as a post-starburst galaxy}

NGC~1569 has been suggested to be in a post-starburst phase (Israel
1988a). The definition of the starburst and the post-starburst phase is a matter
of choice. Here we define the starburst phase as being characterized
by ongoing SF. With this definition the field population of NGC~1569 clearly
is in a post-starburst phase since we find that the CMD requires no
SF activity over the last few Myr. 

On the other hand, in the same
region there are many HII regions and three SSCs, whose age is likely 
to be younger
than 10~Myr (Gonz\'alez-Delgado \etal\ 1997). Therefore  
the stellar population in NGC~1569 is made of
two components: the resolved stars with ages older than $\sim$ 10~Myr, and the
stars in the SSCs, which formed more recently. The resolved stellar component
is in a post-starburst stage, while the most recent star formation is
taking place in the SSCs. This finding may suggest 
a picture of discrete SF events mainly generating SSCs, which dissolve
to create the field stellar population in a certain time interval. 
The $\sim$10 Myr quiescence period would correspond to a
characteristic  
time step of the discrete SF activity. On the other hand,    
the compactness of the SSCs suggests that they are
gravitationally bound systems, and that they are true young globular
clusters (Ho $\&$ Filippenko 1996). In addition, the age of the
brightest component of SSC-A has been estimated to be as old as 8~Myr
(Gonz\'alez-Delgado \etal\ 1997), and still presents a half-light
radius of $\sim$1.6~pc (De Marchi \etal\ 1997). 
Therefore the field population is likely to be the relic of a major
episode of SF which involved the whole observed region, and
which ended several~Myr ago. The SF activity in the field is most likely
responsible for the outflow of the interstellar medium analyzed by Waller
(1991), who derived a kinematic age of 10 to 60~Myr for the H$\alpha$ filaments
around NGC~1569. This SF activity, however, has not 
depleted the gas reservoir, and more recent SF is currently
occurring in the HII regions and in the SSCs. This may suggest that
only the densest regions are able to collapse to form stars
following the burst of SF and its consequent energy input in the ISM.

\subsection{Comparison with other dwarf irregular galaxies}

The SFR in dwarf irregulars has been
found to be characterized by long activity periods, separated by short
interburst phases ({\it gasping SF}, e.g., Tosi 1994). In this respect,
the field stellar population in NGC~1569 exhibits similar properties,
with a SF occurring at an approximately constant
level over $\sim$0.15~Gyr, with quiescence periods shorter than $\sim$10~Myr. 
A quantitative comparison is instructive.
Among the different possible ways in which the strength of the star 
formation activity can be characterized, we consider the SFR per unit
surface (see Hunter $\&$ Gallagher 1986).

At a distance of 2.2 Mpc, the observed PC field surveys an area of 
approximately
0.14~kpc$^2$, so that our derived recent SF rates are 
about 20, 7, and 4~\Myr~kpc$^{-2}$, for IMF slopes of 3, 2.6,
and 2.35, respectively (again having extrapolated the IMF down to
0.1~\Ms). 
It is worth recalling that the typical SFR quoted for the solar
neighborhood (e.g. Timmes, Woosley, $\&$ Weaver 1995) is $\sim$1000 
times lower than this value.

The comparison of the SFR in NGC~1569 with those derived for other dwarf
irregulars is not straightforward since there is a non-trivial
dependence on the method used to measure the SFR. Even when considering
the SFR derived from the analysis of the CMD via theoretical simulations, there
may be systematic differences due to the adopted set of stellar tracks
(Greggio 1994). Finally, the SFR depends on the adopted IMF, both on
the slope and on the mass cut-offs, the latter being not always
clearly specified in the literature. 
   
With this caveat in mind, the recent SFR derived via theoretical
simulations for
dwarf irregular galaxies in the Local Group range from approximately
$10^{-4}$~\Myr~kpc$^{-2}$ to $10^{-2}$~\Myr~kpc$^{-2}$ 
(MTGF; Tolstoy 1996; Gallart \etal\ 1996a). When allowing for the
uncertainties introduced  by the different methods used, we can
conclude that the recent SFR in NGC~1569 is $\sim$ 100 times
larger than in the most active local dwarf irregular. NGC~1569 fully
deserves its classification as a (post)-starburst galaxy, as opposed to normal,
star-forming galaxies.

Could it be that strong SF bursts took place in dwarf irregulars in the
past? The chemical properties, and the gas content of the  most
active of them are not consistent with the extrapolation of
the currently observed activity over a long time. Chemical evolution
models show that several bursts of SF may have occurred in these galaxies,
but separated by long quiescence periods. Strong bursts of SF
could perhaps be accommodated but the duty cycle of this activity has to be
low (Matteucci $\&$ Tosi 1985; Pilyugin 1994; Marconi, Matteucci, $\&$
Tosi 1994).

More direct information on the past SF activity occurred in dwarf
irregulars has been derived for some Local Group objects by analyzing the
stellar population along the giant branches (Gallart \etal\ 1996a;
Aparicio, Gallart, $\&$ Bertelli 1997a,b). 
Although the SFRs derived in this way are less detailed and
reliable than those deduced from the analysis of the MS and core
He-burning stars,
there is no hint that these galaxies experienced a much
stronger SFR in the past than recently. On the other hand, the analysis of
the giant branches does not permit a fair time
resolution, and the past activity in these galaxies could also have
been characterized by recurrent strong bursts, whose average rate is
reflected by the bright
portion of the CMD.

The CMD derived by VB
exhibits a well populated AGB, which definitely argues for 
previous SF in NGC~1569. VB
concluded that a SF episode took place at a significantly lower rate
than the most recent one between 1.5 and 0.15~Gyr ago. However,
the values of these parameters are uncertain due to observational
and theoretical limitations. Nevertheless, the mere presence of AGB
stars gives evidence for a previous SF activity at intermediate ages.
Ages as old as $\sim$10 Gyr can be probed only with NICMOS 
observations looking for the presence of stars at the tip of the first
red giant branch in the infrared.

In the meantime, indirect information on the past
SF history can be derived from chemical
evolution arguments. A recent determination of the oxygen, nitrogen
and helium abundances in
the interstellar medium of NGC~1569 was done by Kobulnicky \& Skillman
(1997). Comparing their values to the predictions of
chemical evolution models for bursting dwarfs  
(Marconi, Matteucci, $\&$ Tosi 1994) we find that 
several bursts of SF
may have happened in this galaxy, provided that strong 
galactic winds have occurred. The products of SNII explosions
should be preferentially lost in these winds, thus keeping the overall
metallicity low.

To summarize, the comparison of
the SF history in NGC~1569 and that of dwarf irregular galaxies
in the Local Group is inconclusive. While the general characteristics of the SF
appear very similar, the level of the current SFR in NGC~1569
seems significantly larger than that of the dwarf irregulars
in the Local Group over their whole lifetime. This suggests
that NGC~1569 is distinctively
more active than the Local Group dwarf irregulars.  

An evolutionary link between dwarf spheroidals and dwarf irregulars
has been suggested several times (e.g., Kormendy 1985; Binggeli 1994).
Strong bursts of SF are known to
have occurred in dwarf spheroidals in the past
(e.g., Smecker-Hane et al. 1994). Unfortunately, to our knowledge
a quantitative determination of the level and duration of
these bursts has not yet been attempted.
The evidence that NGC~1569 is currently experiencing a galactic outflow
seems well documented, an effect which
would turn the galaxy into an inconspicuous, perhaps forever fading object. 
On the other hand, the ultimate fate of the expanding gas is
rather uncertain, due to the complications of hydrodynamic
modelling. A stronger case could perhaps be made by comparing the SFR in
the bursts which occurred in the dwarf spheroidals, to what we measure
for NGC~1569.

\subsection{ Implications for the faint blue galaxies}

We now consider the possible relevance of galaxies like
NGC~1569 to the interpretation of faint galaxy counts. To do this, we
first discuss the possible evolutionary path for the SFR in this
galaxy.

The direct census of the stellar population in NGC~1569 suggests that
a strong burst of SF has occurred at a rate of $\sim$0.5~\Myr, 
with a duration of $\sim$0.15~Gyr.
This level of the SFR cannot have been sustained over a long period of
time. The total mass of the galaxy as estimated by Israel (1988a)
is $\sim$3.3~$\times10^8$~\Ms.
Considering a SFR of 0.5~\Myr\ as representative of the
whole galaxy (i.e. neglecting the activity in the regions external to
our PC field, and considering a turn-over in the IMF at the low mass
end) it only takes $\sim$0.7~Gyr to cycle the entire mass into stars.
The gas exhaustion timescale is longer than this because of the mass
return from stellar winds and supernovae. However, this corresponds to
a small correction since 1~Gyr after its birth 
a single age stellar population has returned
only $\sim$0.2 of its total mass (for a Salpeter IMF). Thus it seems
very unlikely that such a high SFR could have been supported
throughout a lifetime of several Gyr, even allowing for the occurrence of
galactic winds in the past, which would imply a larger initial
mass. Furthermore, the metallicity resulting from such extreme
activity would inevitably be much larger than derived from the spectra
of HII regions. Though the above estimate for the gas 
exhaustion timescale is sensitive to the adopted lower mass cut-off,
the metal production depends on the mass range that we directly observe.   
We thus suggest that, in spite of the rather constant
SFR characterizing the more recent epochs, the average past SFR in NGC~1569 
should have been much lower, perhaps with short intense bursts coupled
with long quiescence periods.

Considering the energy input from the recent burst of SF, VB suggest
that at least part of the gas currently present in
NGC~1569 will be lost from the galaxy. We confirm this expectation, since
our SNII rate is consistent with what found by VB.
However, the burst giving birth to the stars in our observed
field did not completely exhaust the gas reservoir, since the SSCs and HII
regions are younger than the field. It could be that, while the burst
of SF was able to initiate a general outflow of the ISM, the highest density
clumps succeeded in collapsing to form stars, possibly stimulated
by the SN explosions. The ultimate fate of the outflowing gas depends
on several uncertain factors, including dissipation. In addition,
hydrodynamic models for bursting dwarfs suggest that, once
the energy input from stellar winds and supernovae has ceased, the
interstellar medium in the outer disk of the galaxy falls back to the
center, where it could power a successive burst of SF (D'Ercole $\&$
Brighenti 1998). 
 
To summarize, the recent burst could be the strongest SF episode in
this galaxy over its entire lifetime, though the future evolution is
rather uncertain. 
Among the various models proposed in the literature to
explain the excess population of faint blue 
galaxies, we consider those by Gronwall $\&$ Koo (1995) and Babul $\&$
Rees (1992). In the model proposed by Gronwall $\&$ Koo (1995), the
blue galaxy population does not evolve. 
The luminosity evolution of NGC~1569 is likely to 
be characterized by on and off phases, corresponding to
the recurrent bursts, with long interburst periods. 
Therefore, if NGC~1569 were a typical object in the class of galaxies
mostly contributing to the excess blue objects,
a constant luminosity evolution of the population could be
accomplished only with an
appropriate number evolution of the galaxies of this kind --- not a very
attractive scenario. In other words, it seems unlikely that NGC~1569
represents the local counterpart of the galaxies mostly contributing to
the faint blue counts in the Gronwall $\&$ Koo (1995) model.

According to Babul $\&$ Rees (1992), the excess blue objects are
bursting dwarf galaxies (at redshifts close to $\sim$1) which, 
after the burst, become gas-free because of SN explosions. 
Later evolution is bound to passive fading. Thus, 
NGC~1569 cannot be regarded as a possible descendant of
the bursting dwarfs, since it is bursting now. Still, its properties 
may be relevant to the Babul $\&$ Rees model.
In a quantitative study of this models,
Babul $\&$ Ferguson (1996) show that
the current constraints on the faint galaxy population can be met 
if the SFR in the bursting dwarfs is large, so that the object is
bright enough
to contribute to the counts at the appropriate redshift. Our derived
SFR for the most recent burst in NGC~1569 seems close to the rate of 
$\sim$1~\Myr\ required by their model. However, Babul $\&$ Ferguson
adopt a burst duration of 10~Myr, one order of magnitude lower than
what found here for NGC~1569. The success of their model depends on
the combination of the duration of the burst, efficiency of star
formation, and on the decay rate of the dwarf galaxy formation
probability. It is likely that longer burst durations can be
accommodated when varying the other parameters.
The model by Babul $\&$ Ferguson also requires substantial 
dimming of the bursting dwarfs after the major burst. This is needed
in order to avoid an excess of objects at brighter magnitudes, where
the counts are dominated by the evolution of bigger galaxies.
 As emphasized
before, we do not know if NGC~1569 will definitely lose its gas. 
The fact that NGC~1569 has
experienced at least one previous burst and that the most recent one has had
a relatively long duration, raises doubts on this hypothesis.
Nevertheless, the case of NGC~1569 demonstrates that 
dwarf galaxies are capable to sustain bursts of SF of the magnitude
required by this model.

\section{Summary and conclusions}
 
In this paper we have presented HST WFPC2 F555W and F439W photometry of 
resolved stars in the dwarf irregular galaxy NGC~1569. About
2800 stars were measured with photometric error smaller than 
\sigmada = 0.2\,mag out of a total of $\sim$7000 objects detected.
The corresponding CMD extends down to \mb, \mv~$\simeq$~26, and is
complete for \mv~$\lsim$~23. It samples stars more massive than 
$\sim$4~\Ms\, allowing us to derive the SF history over the
last 0.15~Gyr.

The interpretation of the data has been performed via theoretical 
simulations, based on stellar evolutionary tracks. 
We have considered several sets of stellar tracks from the Geneva and
Padova groups, and were able to reach a satisfactory representation
of the data adopting either the Geneva tracks with $Z$~=~0.001,
or the Padova tracks with $Z$~=~0.004. Since the applicability of these tracks
is related to the shape of the blue loops, which is sensitive to
details in the input parameters used in the computations, we consider it
premature to dismiss alternative sets of evolutionary tracks.

The two sets provide a similar picture of the recent SF history in the field
of NGC~1569. The galaxy has experienced a general burst of SF, of
duration $\Delta t \gsim 0.1$~Gyr, at an approximately constant rate.
The burst seems to have stopped $\sim$5~--~10~Myr ago but SF continues
in the HII regions and in the SSCs.
If this general burst consists of successive episodes, these must have
occurred at a similar rate, and be separated by short quiescent
periods. Qualitatively, this behavior looks similar to that inferred for
dwarf irregulars in the Local Group. Quantitatively, though, the level
of the SFR in this recent burst in NGC~1569 is approximately 2 orders
of magnitude higher. 

We find that  the Salpeter IMF is 
consistent with the observed CMD and LF, but our best simulations are
characterized by slightly steeper slopes (2.6 -- 3). This seems to
disagree with observations of other starbursts which generally indicate
a Salpeter slope (Leitherer 1997). The present study, however, refers to the
field population of stars less massive than $\sim$30~\Ms, whereas most 
starburst IMFs are derived for the nuclear
burst and for dense OB clusters in the mass range above $\sim$20~\Ms.
Since most of the SF activity in the NGC~1569 field has subsided about 
7~--~10~Myr ago, we have no direct information on the IMF of those stars with
lifetimes less than this value. The corresponding minimum masses are around
30~\Ms. The field-star IMF in our Galaxy and in the
Magellanic Clouds is steeper (biased against massive stars)
than the cluster IMF (Massey, Johnson, $\&$ DeGioia-Eastwood 1995a;
Massey \etal\ 1995b), with the cluster value being
close to Salpeter. This could be due to a richness effect, where the most
massive stars are not observed simply because of their small expected numbers.
NGC~1569 may be another case where the most massive stars
with masses above $\sim$30~\Ms\ are preferentially formed in clusters, but
we emphasize the different mass range sampled by our and that of LMC/SMC
studies. Our preferred IMF is quite similar to that derived
for field stars of comparable mass in our Galaxy. This is in agreement with the
notion of a virtually constant IMF, at least as far as the shape is concerned,
and does not support the expectations of flatter slopes in
starbursting regions (e.g., Padoan, Nordlund, $\&$ Jones 1997), or in
low metallicity environments, as sometimes invoked to better reproduce
the properties of elliptical galaxies (e.g., Vazdekis \etal\ 1996).

In the past, the SF in NGC~1569 is likely to have proceeded at a
substantially lower average rate than in the recent burst. This follows from
the estimate of the gas exhaustion timescale, and from the relatively
low metallicity of the ISM in this galaxy. This leaves the
possibilities of either an approximately constant SFR,
or short strong bursts, with long interburst
periods.
Detailed chemical evolution models are
needed to explore quantitatively the possible evolutionary paths for
NGC~1569, and this will be the subject of a forthcoming paper. 
 
The recent burst of SF is certainly not the first episode in NGC~1569 
(see also VB) but may be the last. The velocity field of the ionized
gas (Tomita, Ohta, $\&$ Saito 1994) and energetics arguments
support this possibility (Heckman \etal\ 1995). If this was
the case, in the future NGC~1569 may turn into a dwarf spheroidal
galaxy. In addition, its luminosity evolution would be characterized by
passive fading, which seems required by the Babul $\&$ Ferguson (1996)
bursting dwarfs model to explain the excess faint galaxy counts.
However, it is also possible that the outflowing gas falls back onto the
galaxy, triggering a successive burst of SF, an option which has to be
explored with the computation of detailed hydrodynamic modelling for
bursting dwarf galaxies (D'Ercole $\&$ Brighenti 1998). In spite of
these uncertainties, our results 
shows that dwarf galaxies are capable of sustaining bursts of SF at large
enough rates to be relevant for the interpretation of the faint galaxy
counts.

\acknowledgments

This work has been partially supported by ASI through grant 9770070AS.
Support for this work was also provided by NASA grant GO-06111.01-94A 
from the Space Telescope Science Institute, which is operated
by the Association of Universities for Research in Astronomy, Inc., under
NASA contract NAS5-26555. L.G. and M.T. gratefully acknowledge the 
hospitality of STScI.

\clearpage

\figcaption{Photometric errors versus magnitude as given by DAOPHOT 
in the F439W (left panel) and F555W (right panel) filters.
\label{fig-sigma}}

\figcaption{CMD of the stars detected in the PC field of NGC~1569: a) objects 
with photometric error $<$0.2\,mag, b) objects with error $<$0.1\,mag.
\label{fig-cmdo}}

\figcaption{ Luminosity function of stars with photometric error
\sigmada $<$0.1.
Full dots correspond to the whole set of data shown in
Fig.~\ref{fig-cmdo} (panel b),
open circles to the {\it reference blue plume} defined in the text.
\label{fig-fl}}

\figcaption{Geneva tracks with $Z=0.001$ in the observational plane,
having adopted $E(B-V)=0.56$, $(m-M)_0=26.71$. The faintest loop
appearing in the plot belongs to the 5~\Ms track. The other sequences
have $M$=7,9,12,15,20,25,40,60,85,120~\Ms.  Only the main-sequence evolution 
is plotted for the three most massive sequences. Each dot corresponds
to a model in the evolutionary sequence. 
The slow evolutionary phases can be easily recognized from the high
density of dots along each track.  
\label{fig-cmdth001}}

\figcaption{ Synthetic CMD computed with Geneva tracks.
Panel a) shows one of the best cases obtained  with the 
$Z=0.008$ set; panel b) shows one of the best results with the
$Z=0.004$ set. The corresponding global LFs 
(solid line for the $Z=0.008$ case, dotted  for the $Z=0.004$ case) is
compared with the observational LF in panel c). See text for
more details.
\label{fig-synth1fl}}

\figcaption{Geneva tracks with $Z=0.008$ in the observational plane,
having adopted $E(B-V)=0.46$, $(m-M)_0=26.71$. The faintest loop
appearing in the plot belongs to the 5~\Ms\ track. The other sequences
have $M$=7,10,12,15,20,25,40,60,85,120~\Ms. For the three most massive
sequences only the Main Sequence evolution is plotted. Each dot corresponds
to a model in the evolutionary sequence. 
\label{fig-cmdth008}}

\figcaption{Synthetic CMD computed with Geneva tracks with $Z=0.001$. Panel a) 
shows the best result obtained with constant SF, 
panel b) shows the best result 
obtained with an exponentially decreasing SF. Both simulations assume
no SF activity during the last 7 Myr. The corresponding global LF 
(solid line for the panel a model, dotted for the panel b model) is
compared with the observational LF in panel c.
\label{fig-synth2fl}}

\figcaption{Synthetic CMD computed with Geneva tracks with $Z=0.001$. 
Panels a) and b) show the best results obtained with two distinct 
episodes of SF, the older one ending 0.05 Gyr ago. Compared to the
case shown in panel b), in the
simulation shown in panel a) we have adopted a larger
SFR in the most recent episode. This trend is counterbalanced
by the different IMF exponent. In panel c)
we show the luminosity functions for the models in panel a) (solid
line) and panel b) (dotted line), and the observed LF (dots).
\label{fig-synth3fl}}

\figcaption{Synthetic CMD computed with the Padova tracks with Z=0.004. 
Panels a) and b) show the best results obtained with two episodes of
SF, the older one ending 0.035 Gyr ago. In panel c)
we show the luminosity functions for the models in panel a) (solid
line) and panel b) (dotted line). In both cases the SFR in the
most recent episode is slightly lower than in the previous one, and
one can see the effect of varying the IMF exponent: the steeper slope
(panel b)) tends to underpopulate the brightest bins of the LF. No SF
is assumed to have occurred in the last 10 Myr. 
\label{fig-synth4fl}}

\figcaption{Synthetic CMD computed with the Geneva tracks with
Z=0.001, a constant SF and $\alpha$=3,
but adopting different epochs for the end of the burst: currently
ongoing SF (panel
a), stopped at 5 \Myr ago (panel b). The corresponding global LF 
(solid line for the panel a) model, dotted for the panel b) model) is
compared with the observational LF in panel c).
\label{fig-synth10fl}}

\figcaption{Synthetic CMD computed with the Padova tracks with Z=0.004,
$(m-M)_0$ = 26.71 and $E(B-V)$=0.46.
The synthetic CMD in panel a) has been obtained assuming two episodes
of SF, separated by a quiescence
period of 10 Myr (from 40 to 30 Myr ago), and no SF in the last 10
Myr. In the simulation in panel b) a constant, currently ongoing SF
has been adopted. The corresponding LFs (solid line for the CMD in the top
panel) are shown in panel c).
\label{fig-synth6fl}}

\figcaption{Synthetic CMD computed with the Padova tracks with Z=0.004,
and adopting $(m-M)_0$ = 28.01 and $E(B-V)$=0.46.
The simulation in panel a) has been computed assuming two episodes of
SF, the most recent one stopped 5 Myr ago. In the simulation in panel b)  
a currently ongoing, constant SFR has been adopted. As usual, panel c)
shows the corresponding LFs (solid line for the CMD in the top panel).
\label{fig-synth8fl}}

\clearpage

\begin{table}
\caption{\label{tab-compl} Photometric completeness as
a function of 
the magnitude in the F439W band}
\begin{center}
\begin{tabular}{cc}
\hline\hline
\multicolumn{1}{c}{$m_{439}$} &
\multicolumn{1}{c}{completeness} \\
\hline
$ < 23.0 $	&  $ > 99\%$		\\	
$23.0-24.0$	&  $ \simeq  80\%$	\\  
$24.0-25.0$	&  $ \simeq  50\%$  	\\
$25.0-26.0$	&  $ \simeq  30\%$	\\
\hline
\end{tabular}
\end{center}
\end{table}

\clearpage

\begin{table}
\caption{\label{tab-mod} Stellar evolutionary tracks adopted for the 
synthetic CMDs}
\begin{center}
\begin{tabular}{lccc}
\hline\hline
Model & $Z$ & $\dot{M}$ & Reference \\
\hline
Geneva & 0.008 & standard & Schaerer \etal\ (1993) \\ 
Geneva & 0.008 & high & Meynet \etal\ (1993) \\ 
Geneva & 0.004 & standard & Charbonnel \etal\ (1993) \\ 
Geneva & 0.004 & high & Meynet \etal\ (1993) \\ 
Geneva & 0.001 & standard & Schaller \etal\ (1992) \\ 
Padova & 0.004 & standard & Fagotto \etal\ (1994) \\
\hline
\end{tabular}
\end{center}
\end{table}

\clearpage

\begin{figure}
\epsscale{1.1}
\psfig{figure=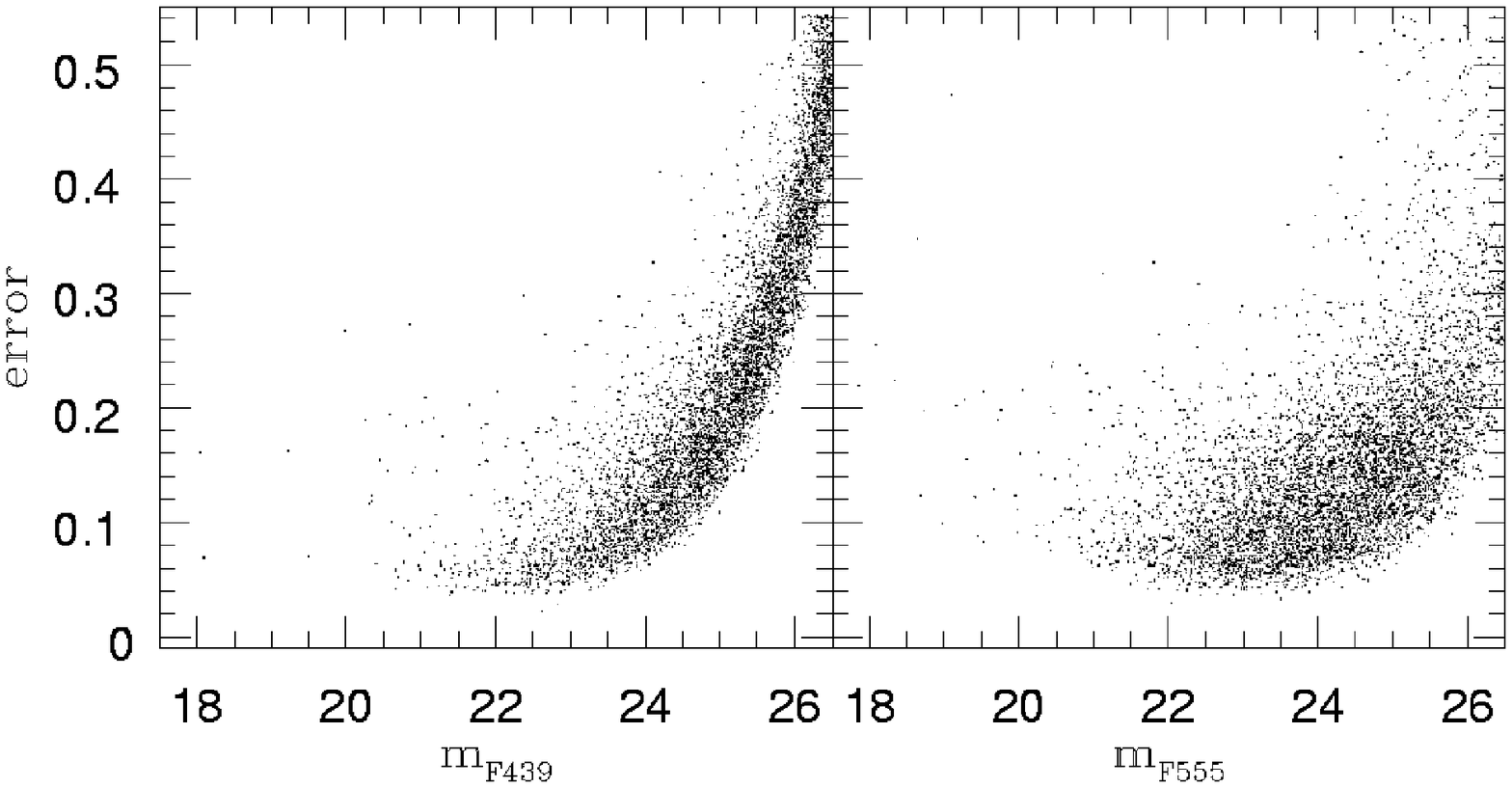,width=15cm}
\end{figure}

\begin{figure}
\epsscale{1.1}
\psfig{figure=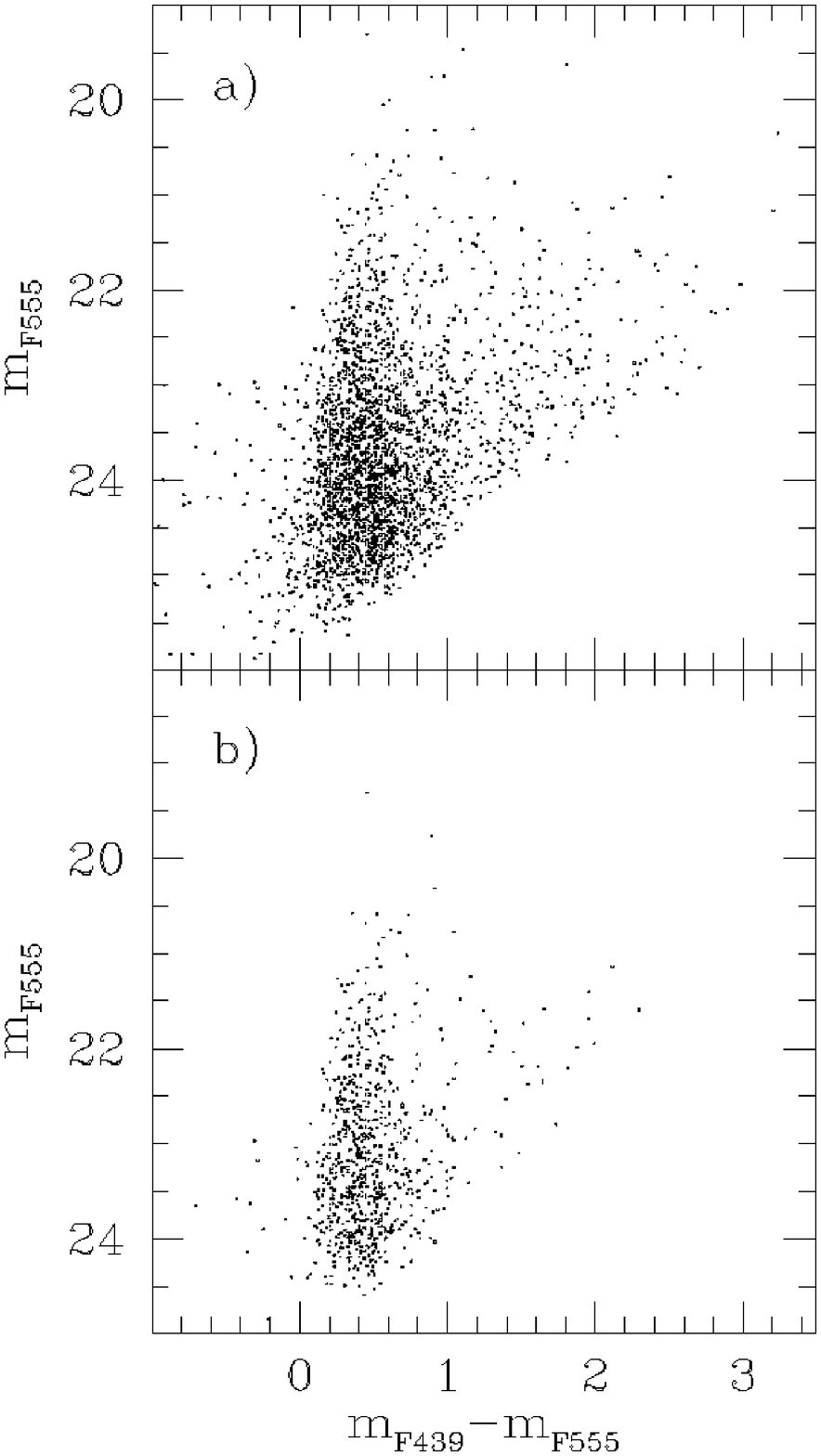,width=10cm}
\end{figure}

\begin{figure}
\epsscale{1.1}
\plotone{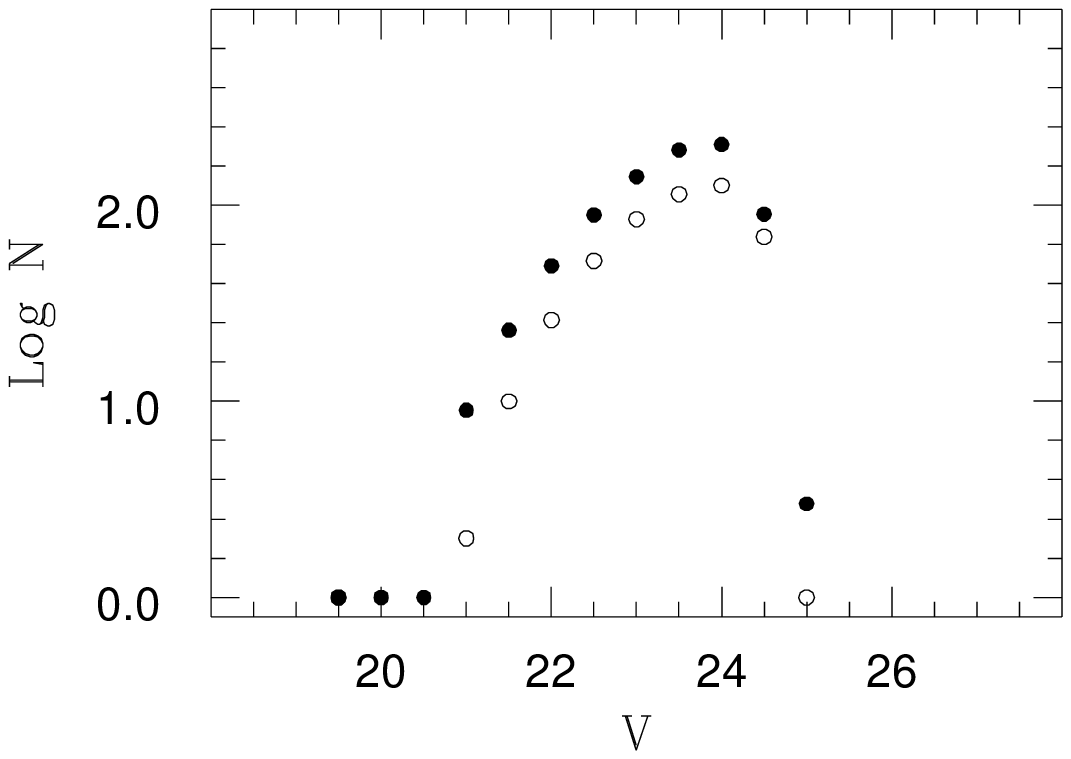}
\end{figure}

\begin{figure}
\epsscale{1.1}
\plotone{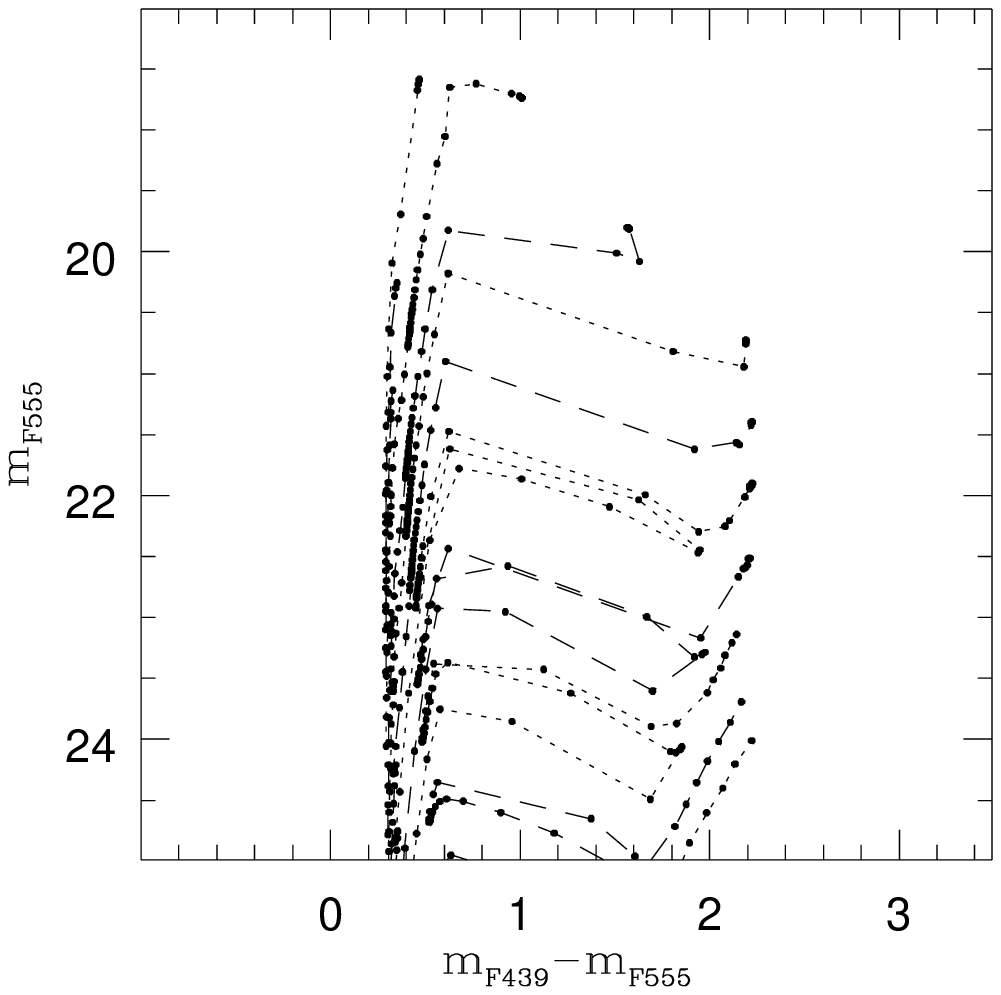}
\end{figure}

\begin{figure}
\epsscale{1.1}
\plotone{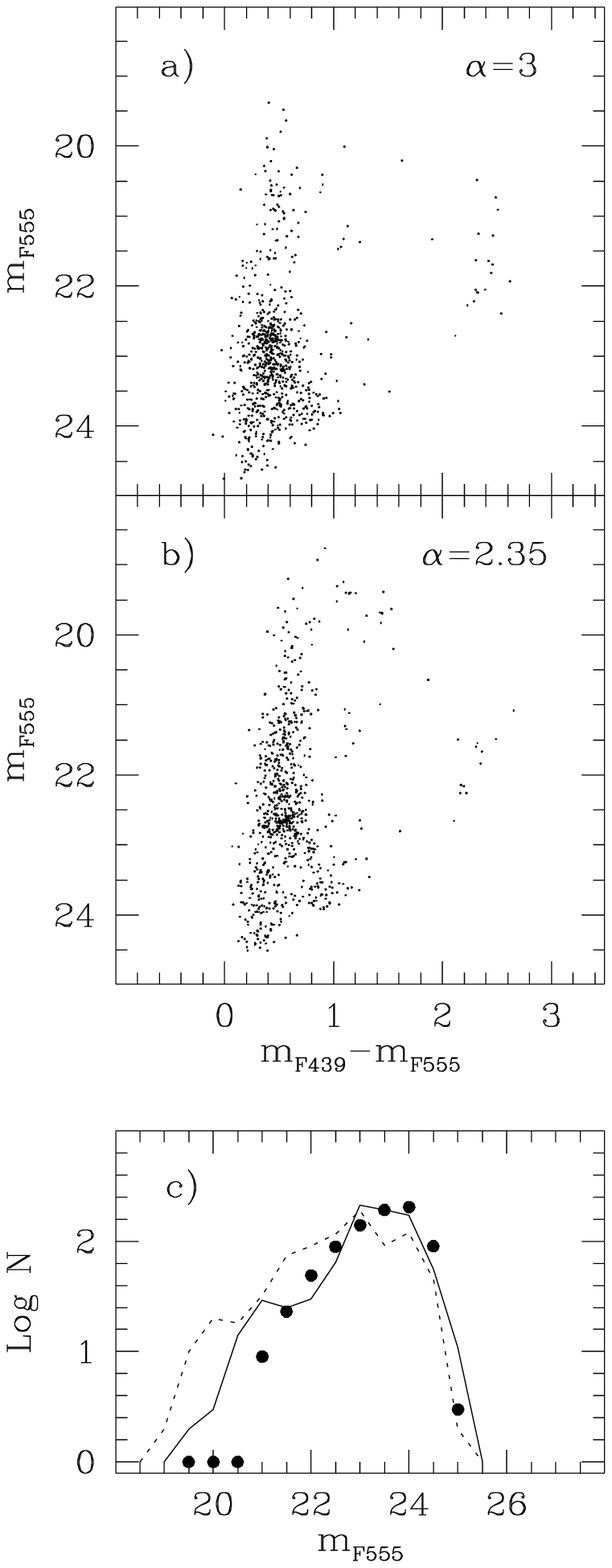}
\end{figure}

\begin{figure}
\epsscale{1.1}
\plotone{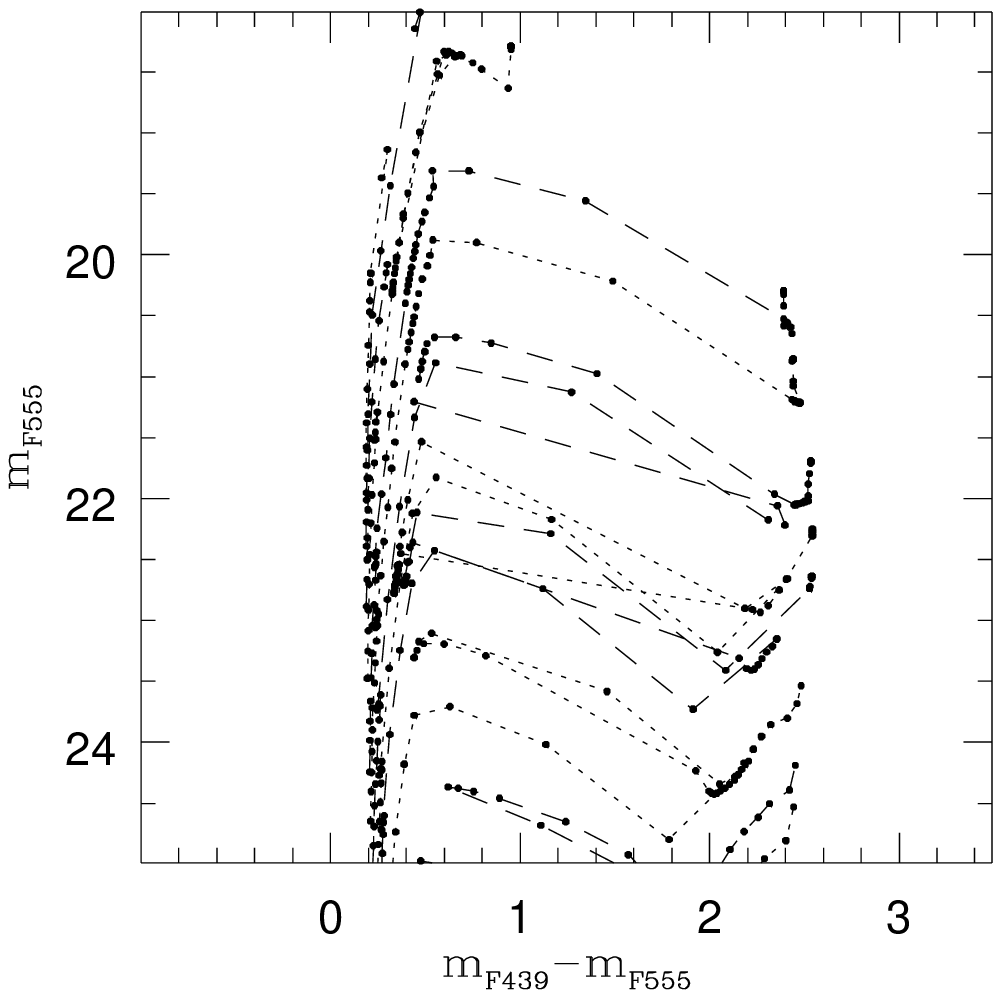}
\end{figure}

\begin{figure}
\epsscale{1.1}
\plotone{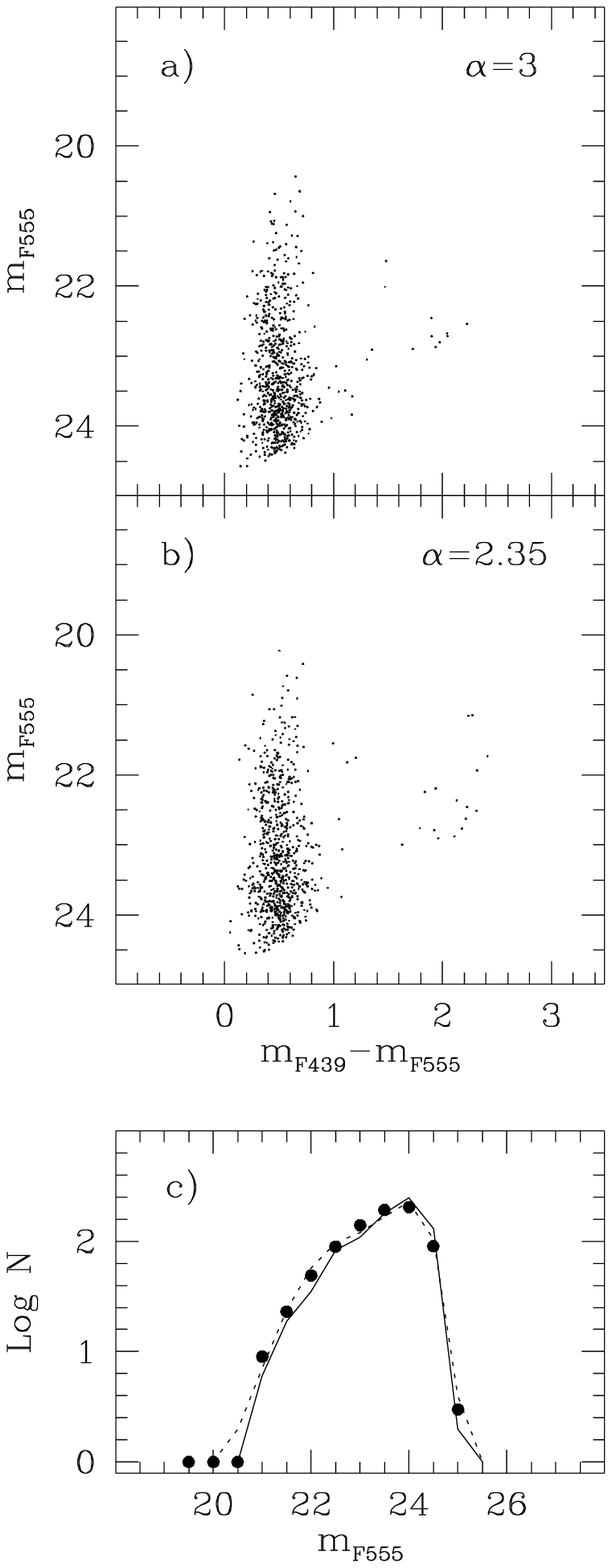}
\end{figure}

\begin{figure}
\epsscale{1.1}
\plotone{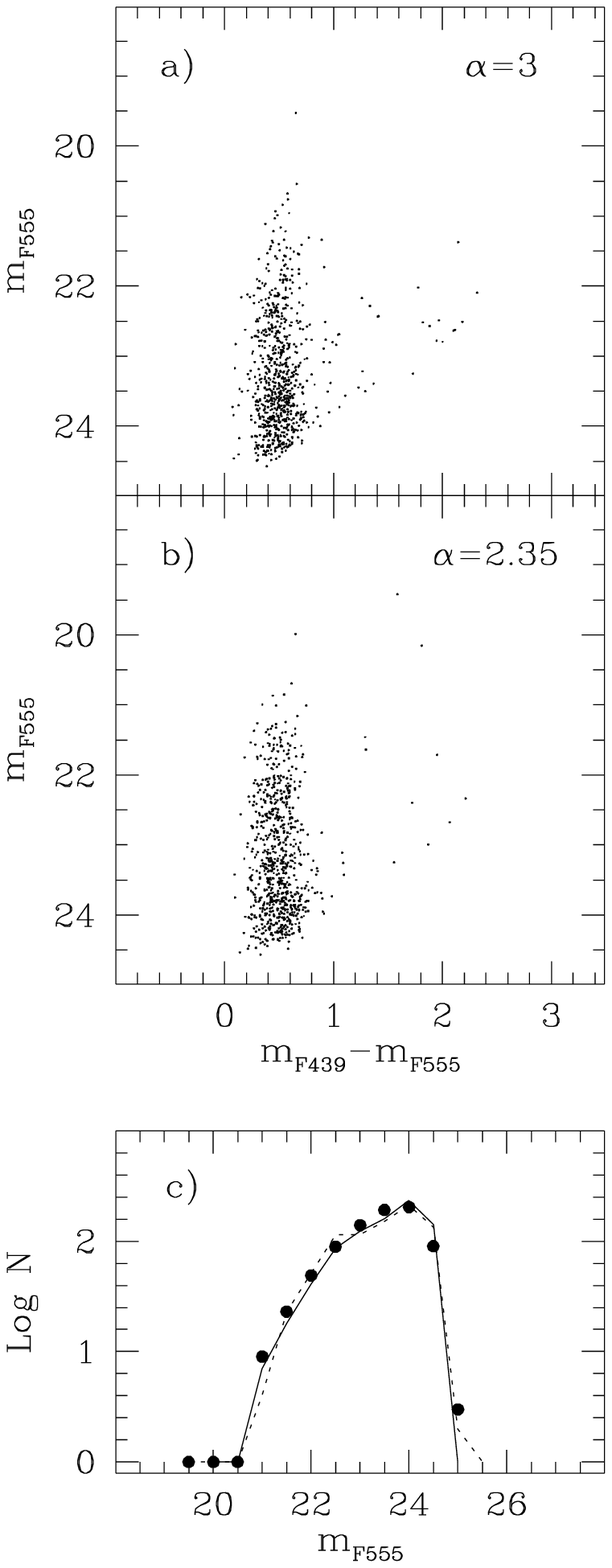}
\end{figure}

\begin{figure}
\epsscale{1.1}
\plotone{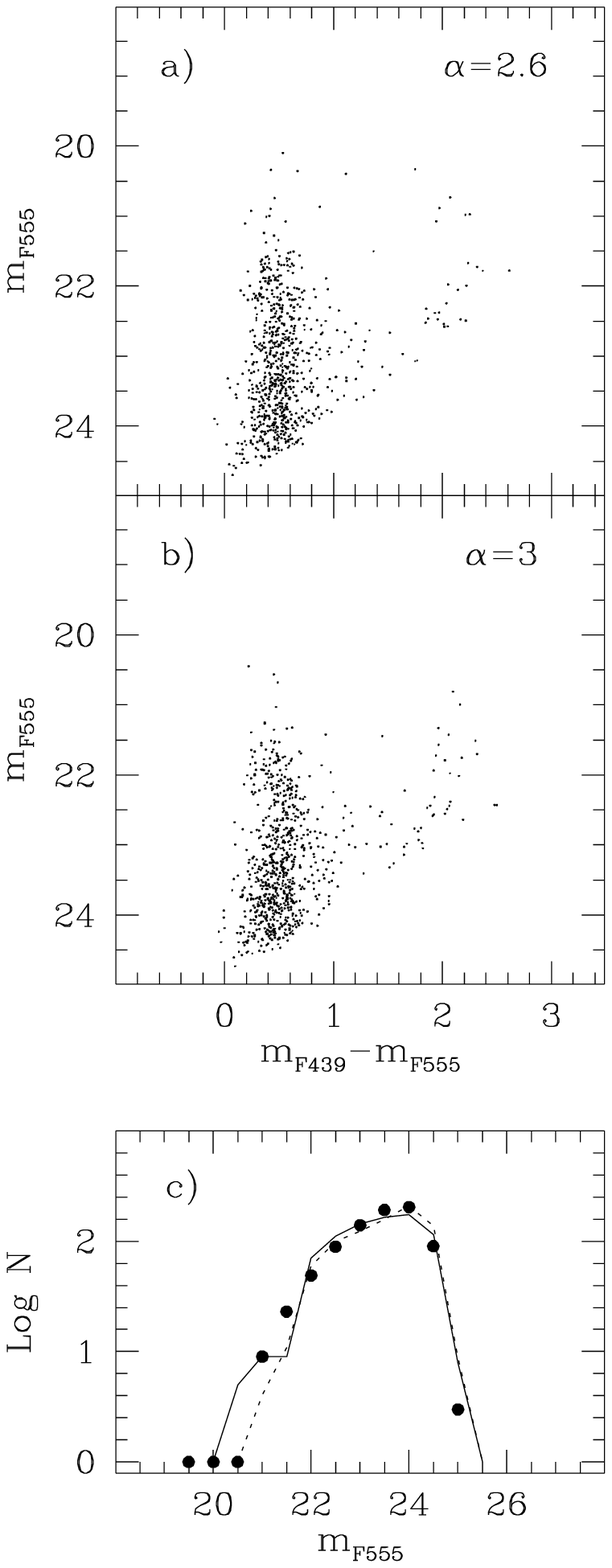}
\end{figure}

\begin{figure}
\epsscale{1.1}
\plotone{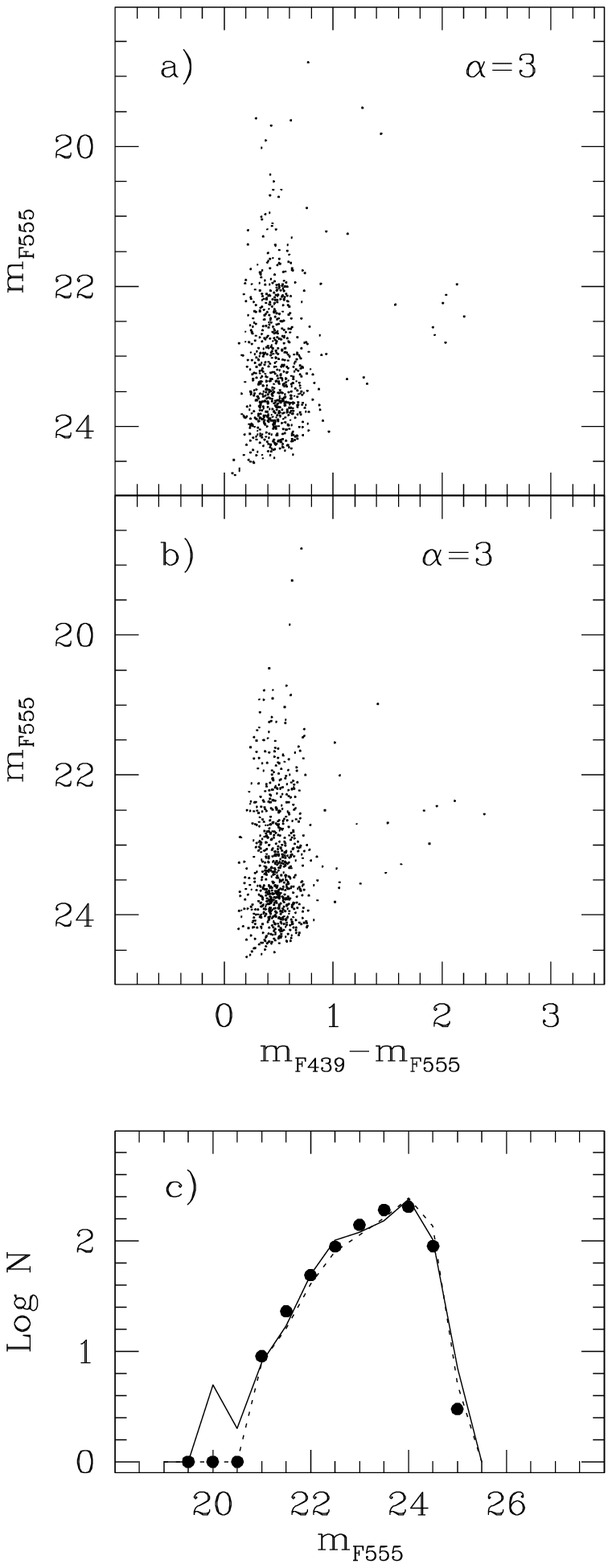}
\end{figure}

\begin{figure}
\epsscale{1.1}
\plotone{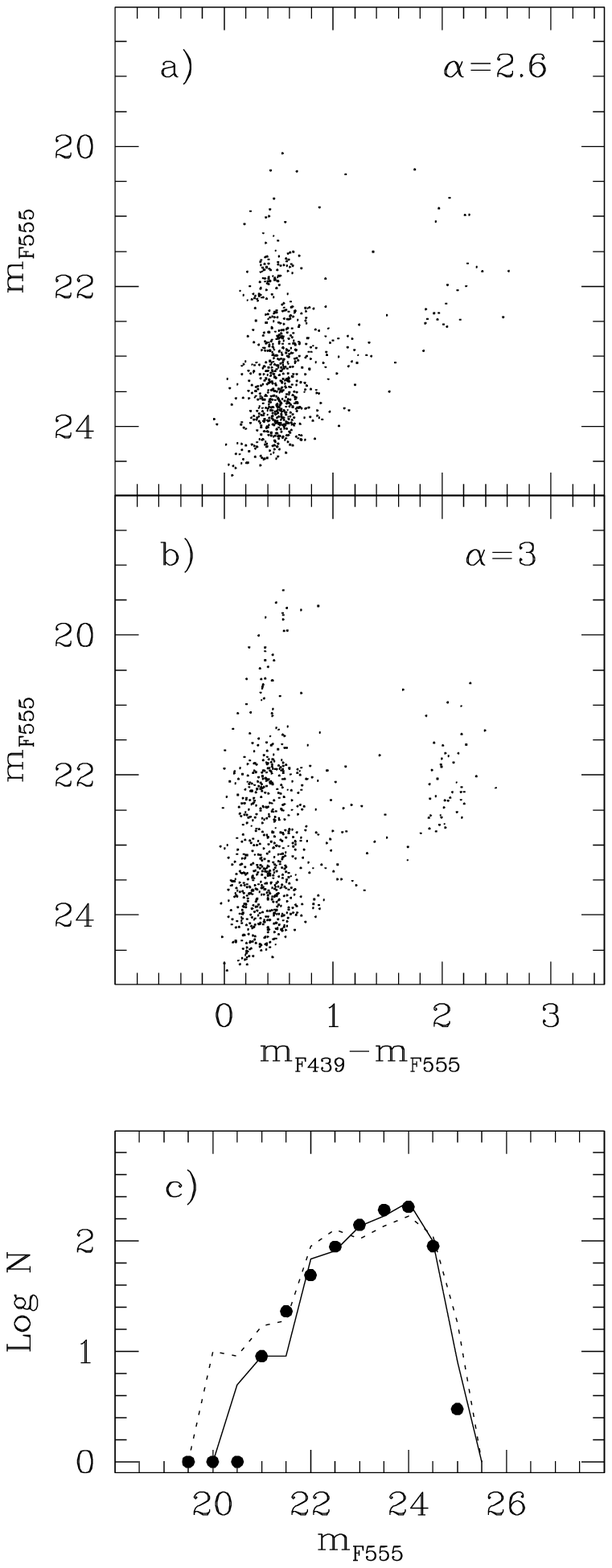}
\end{figure}

\begin{figure}
\epsscale{1.1}
\plotone{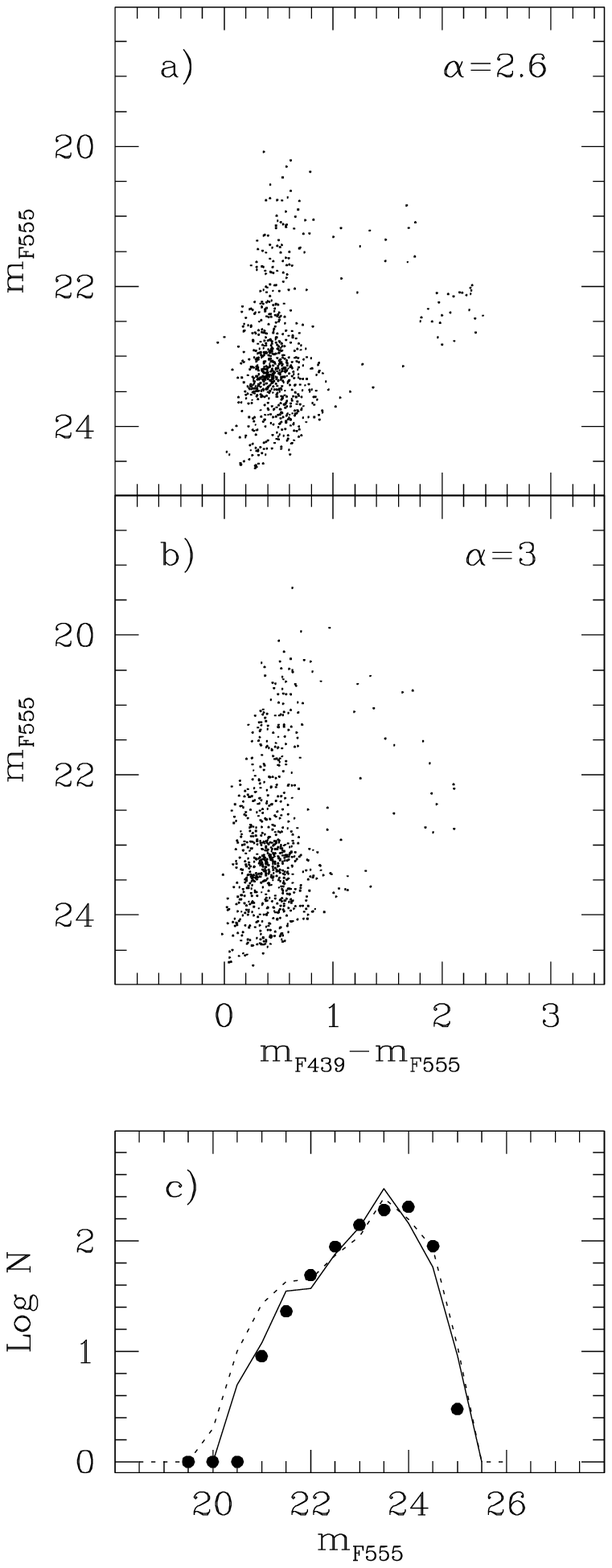}
\end{figure}

\end{document}